\documentclass{article}
\usepackage{arxiv}

\usepackage[utf8]{inputenc}
\usepackage[colorlinks,bookmarksopen,bookmarksnumbered,citecolor=red,urlcolor=red]{hyperref}
\hypersetup{final}

\usepackage{appendix}

\usepackage{bookmark}
\usepackage{soul}
\usepackage{amsthm}
\usepackage{amsmath}
\usepackage{amsfonts}
\usepackage{listings}
\usepackage{float}
\usepackage{caption}
\usepackage{wrapfig}
\usepackage{subcaption}
\usepackage{titlesec}

\usepackage{mathtools,cuted}
\everymath{\displaystyle}

\titlespacing\section{0pt}{12pt plus 4pt minus 2pt}{0pt plus 2pt minus 2pt}
\titlespacing\subsection{0pt}{12pt plus 4pt minus 2pt}{0pt plus 2pt minus 2pt}
\titlespacing\subsubsection{0pt}{12pt plus 4pt minus 2pt}{0pt plus 2pt minus 2pt}

\usepackage{xargs}                      

\usepackage{enumitem}
\SetLabelAlign{parright}{\strut\smash{\parbox[t]{\labelwidth}{\raggedleft#1}}}

\usepackage[dvipsnames]{xcolor}
\usepackage{multicol}

\usepackage{moreverb,url}
\usepackage{mathtools}

\newtheorem{theorem}{Theorem}

\newtheorem{definition}{Definition}

\newcommand{\dint}{\delta_{int}}
\newcommand{\dext}{\delta_{ext}}

\newcommand\BibTeX{{\rmfamily B\kern-.05em \textsc{i\kern-.025em b}\kern-.08em
T\kern-.1667em\lower.7ex\hbox{E}\kern-.125emX}}

\usepackage[colorinlistoftodos,prependcaption,textsize=tiny]{todonotes}

\newcommandx{\unsure}[2][1=]{\todo[size=\normalsize,linecolor=red,inline,backgroundcolor=red!25,bordercolor=red,#1]{#2}}
\newcommandx{\change}[2][1=]{\todo[size=\normalsize,linecolor=blue,inline,backgroundcolor=blue!25,bordercolor=blue,#1]{#2}}
\newcommandx{\info}[2][1=]{\todo[size=\normalsize,inline,linecolor=OliveGreen,backgroundcolor=OliveGreen!25,bordercolor=OliveGreen,#1]{#2}}
\newcommandx{\wip}[2][1=]{\todo[size=\normalsize,linecolor=lightgray,inline,backgroundcolor=Plum!25,bordercolor=Plum,#1]{#2}}
\newcommandx{\thiswillnotshow}[2][1=]{\todo[disable,#1]{#2}}

\newcommand*{\fullref}[1]{\hyperref[{#1}]{\autoref{#1}}} 

\lstdefinestyle{base}{
    basicstyle=\ttfamily,
 escapeinside={||},
 mathescape=true,
 moredelim=**[is][\color{red}]{@}{@}, 
    basicstyle=\small\ttfamily,columns=fullflexible,
    escapeinside={||},
    mathescape=true
}

\title{EB-DEVS: A Formal Framework for Modeling and Simulation of Emergent Behavior in Dynamic Complex Systems}

\author{
  Daniel J. Foguelman\thanks{Corresponding author}\\
  CONICET-Universidad de Buenos Aires\\
  Instituto de Investigación en Ciencias de la Computación (ICC)\\
  Universidad de Buenos Aires\\
  Facultad de Ciencias Exactas y Naturales\\
  Departamento de Computación\\
  Intendente Güiraldes 2160, C1428EGA Buenos Aires, Argentina.\\
  \texttt{dfoguelman@dc.uba.ar} \\
  \And
  Philipp Henning \\
  University of Rostock\\
  Albert-Einstein-Str. 22, 18059 Rostock, Rostock, Germany \\
  \texttt{philipp.henning@uni-rostock.de} \\
  \And
  Adelinde Uhrmacher \\
  University of Rostock\\
  Albert-Einstein-Str. 22, 18059 Rostock, Rostock, Germany \\
  \texttt{adelinde.uhrmacher@uni-rostock.de} \\
  \And 
  Rodrigo Castro \\
  CONICET-Universidad de Buenos Aires\\
  Instituto de Investigación en Ciencias de la Computación (ICC)\\
  Universidad de Buenos Aires\\
  Facultad de Ciencias Exactas y Naturales\\
  Departamento de Computación\\
  Intendente Güiraldes 2160, C1428EGA Buenos Aires, Argentina.\\
  \texttt{rcastro@dc.uba.ar} \\
}

\begin{document}
\maketitle






 


\begin{abstract}
Emergent behavior is a key feature defining a system under study as a complex system. Simulation has been recognized as the only way to deal with the study of the emergency of properties (at a macroscopic level) among groups of system components (at a microscopic level), for the manifestations of emergent structures cannot be deduced from analysing components in isolation. A systems-oriented generalisation must consider the presence of feedback loops (micro components react to macro properties), interaction among components of different classes (modular composition) and layered interaction of subsystems operating at different spatio-temporal scales (hierarchical organisation).
In this work we introduce Emergent Behavior-DEVS (EB-DEVS) a Modeling and Simulation (M\&S) formalism that permits reasoning about complex systems where emergent behavior is placed at the forefront of the analysis activity. EB-DEVS builds on the DEVS formalism, adding  upward/downward communication channels to well-established capabilities for modular and hierarchical M\&S of heterogeneous multi-formalism systems. EB-DEVS takes a minimalist stance on expressiveness, introducing a small set of extensions on Classic DEVS that can cope with emergent behavior, and making both formalisms interoperable (the modeler decides which subsystems deserve to be expressed via micro-macro dynamics). We present three case studies: flocks of birds with learning, population epidemics with vaccination and sub-cellular dynamics with homeostasis, through which we showcase how EB-DEVS performs by placing emergent properties at the center of the M\&S process.
\end{abstract}


\section{Introduction}
\label{sec:intro}


Complex systems are collections of dynamic components that, upon interaction, may produce novel system-level properties that cannot be directly explained from laws governing the components in isolation~\cite{bar2002general}. 
Thus, complex systems and emergent properties appear tightly connected. 
Sometimes, the very existence of emergent properties is used to \textit{define} a system as being a complex one \cite{Holland2006}. 
System-level properties and interacting components imply at least two organizational levels to be involved in the emergent behavior of these complex systems, i.e., a macroscopic and microscopic level. Thereby, the role of macroscopic level is merely to be the level where the emergent behavior manifests itself. 

This definition assumes varied forms across the literature. For instance in the Springer Complexity program \cite{Springer2020SpringerProgram} it is stated that complex systems are 
\textit{``[\ldots] systems that comprise many interacting parts with the ability to generate a new quality of macroscopic collective behavior the manifestations of which are the spontaneous formation of distinctive temporal, spatial or functional structures.''} 


Others present unexpected properties that arise from low level interactions as non-linear dynamics that yield levels of organization, or even as the formation of order by means of self-organization \cite{Darley1994,Emmeche2000,Kubik2003,holland1998emergence}.  

Thus, emergent behavior shows itself at the macro level of a system, although originating in its micro level, the observed changes are significant and occur spontaneously. 

As a simple illustrative example of emergence in physics we can consider the sand pile. We slowly pour sand over a flat surface. As grains pile up a new structure arises. The novel (and evolving) cone shaped structure confronts the falling grains with a new environment. The pile constraints the degrees of freedom of current incoming grains in ways that differ noticeably from the restrictions faced by the initial batch of grains. In this particular case, moreover, the new macro-structure is \textit{formed by} the micro-constituents themselves, and the system can exhibit self-organized criticality (power-law like distributions of avalanche sizes and duration) \cite{bak1988self}.

Even in this idealized toy system we can identify interesting modeling challenges. We can only know about the existence of the entity ``conical pile'' after we have experimented with, and observed, the system at work. We can not deduce the pile from the laws of motion that govern each individual free-falling grain. Also, the criterion to state when and why a pile has emerged, distinguishing it from simple scattered grains, can be elusive (we can think of the existence of a state-driven structure). Similarly, the \textit{function} of a growing pile (as perceived by grains) will remain the same only until the height of the pile reaches the level of the pouring source (e.g.\ a conveyor belt). 

It is also clear that the spatiotemporal scale at which we decide to analyze the system can in turn modify what we consider an emergent structure (e.g.\ we are ignoring the size, shape, plasticity or rolling capabilities of the grains which may strongly impact, or even prevent, the emergence of a pile).


The above considerations have been dealt with extensively in the literature of complex systems. When it comes to the approach taken by the simulation modeling community, many efforts have been made regarding the identification and validation of emergent properties ex post (after a simulation is completed), while fewer antecedents deal with the live system (at simulation time). In this work we shall provide a formal, system-theoretic modeling and simulation approach to deal with emergence in the live system, relying on the Discrete Event System Specification (DEVS)  \cite{Zeigler:2018:CUC:3213187.3213194}.


When emergence is present as a \textit{concept} already during the modeling phase, it allows the modeler to reason about emergent structures as she encodes the behavior of the system's constituents, explicitly  relating them to the conceived emerging structure. 

Therefore, in this work we will explore how emergence can be integrated into a formal modeling approach, thereby exploring the relations between emergence and multi-level modeling and simulation. 

From a \textit{modeling} perspective placing emergent behavior as a first class citizen will require preserving carefully the separation between macroscopic dynamics, microscopic dynamics and their interaction structure. 
From a \textit{simulation} perspective, emergent properties can appear spontaneously, producing  significant changes within the behavior of a system. Therefore, a hierarchical discrete event modeling and simulation approach will form the basis of our research.  

%

We argue that \textit{understanding} the root causes of a phenomenon and \textit{predicting} its behavior under different conditions can be considered as two sides of the same coin. Yet, it has been argued that in systems with emergence, even having perfect knowledge and understanding may not imply good predictive capabilities, being simulation the optimal means for the study of such systems \cite{Darley1994}. 
This is a relevant issue to deal with when unexpected behavior shows up in engineered systems \cite{Fromm2006} It involves modeling, predicting and analyzing such emergent  behaviors. It is therefore both a concern and a goal to provide sound modeling and simulation technologies capable of producing complex system-level behavior based on individual agent-level behavior, like in self-organizing systems.  


From a multi-level perspective, Wilson states \cite{Wilson1988} that 
\textit{``explanation of observed behavior is not possible with reference solely to the spatial-temporal scale at which the observation was made'' (p.267)}.
           
This requires, in our perspective, working towards modeling and simulation techniques that take into account emergence in an explicit multi-level framework. Consequently, in order to integrate different system levels, communication and causation mechanisms must be established between micro and macro dynamics and structures. We will resort to these methods to share emergent states and to trigger changes in a multi-level setting. 

In this work we present Emergent Behavior DEVS (EB-DEVS), a new approach to tackle the above mentioned issues by relying on the DEVS formal modeling and simulation technique. We will present the formalism and its core theoretical properties such as closure under coupling, bisimulation with Classic DEVS and legitimacy.

Formal approaches facilitate the interpretation and reuse of simulation models by means of clear unambiguous semantics. DEVS is a modeling formalism for discrete-event systems capable of representing exactly any discrete system, and of approximating continuous systems with any desired accuracy. DEVS also makes emphasis on modular and hierarchical composition of (possibly heterogeneous) subsystems. 
In addition, its clear separation of concerns between model definition and model execution will allow us to focus strictly on modeling aspects first, leaving the intricacies of executing a discrete event simulation model as a separate, though very important second stage. 

All these features combined makes DEVS a suitable starting point for our research on modeling and simulating emergent behavior in complex systems. This has been recognised recently as a challenge in the realm of simulation infrastructures for Complex Adaptive Systems where traditional systems engineering practices fall short in capturing emergent behavior  \cite{diallo2018research}\cite{mittal2017simulation}.

The rest of the paper is structured as follows. In \autoref{sec:back} we review the DEVS formalism. Then  
in \autoref{sec:extending} we present the motivations that lead to the new EB-DEVS extension and in \autoref{sec:ebdevs} we provide both the intuitive idea and the formal specification of EB-DEVS, including its theoretical properties. In \autoref{sec:cases} we develop three case studies used to test the capabilities and limits of EB-DEVS as a practical modeling tool. We close the paper in \autoref{sec:discussion} with a discussion about our contribution in relation to other modeling alternatives for complex systems and provide ideas for future research.

\section{Background}
\label{sec:back}

The DEVS formalism~\cite{Zeigler2018TheoryFoundations} can describe hybrid systems, combining discrete time, discrete event and continuous systems. This generality is achieved by the ability of DEVS to represent \textit{generalized} discrete event systems,
i.e., any system whose input/output behavior can be described by sequences of events.

More specifically, a DEVS model processes an input event trajectory and, according to that trajectory and to its own initial state, produces an output event trajectory. Formally, a DEVS \emph{atomic} model is defined by the following structure:
\begin{eqnarray*}
M= \: <X,Y,S,\dint,\dext,\lambda,ta>
\end{eqnarray*}
where:
\begin{description}[noitemsep]
\item $X$ is the set of input event values, i.e., the set of all
    the values that an input event can take.
\item $Y$ is the set of output event values.
\item $S$ is the set of state values. 
\item $\dint$, $\dext$, $\lambda$ and $ta$ are functions which define
    the system dynamics.
\end{description}
Each possible state $s$ ($s\in S$) has an associated \emph{time
  advance} calculated by the \emph{time advance function} $ta(s)$
($ta(s):S \rightarrow \Re^+_0$). The \emph{time advance} is a
non-negative real number representing how long the system remains in a given
state in the absence of input events.

Thus, if the state adopts the value $s_1$ at time $t_1$, after
$ta(s_1)$ units of time (i.e.\ at time $ta(s_1)+t_1$) the system
performs an \emph{internal transition}, resulting in a new state $s_2$.
The new state is calculated as $s_2=\dint(s_1)$, where $\dint$
($\dint:S\rightarrow S$) is called \emph{internal transition
  function}.

Before this state transition from $s_1$ to $s_2$ an output event is produced
with value $y_1=\lambda (s_1)$, where $\lambda$ $(\lambda:S
\rightarrow Y)$ is called \emph{output function}.  

When an input event arrives, the external state transition function $\dext$ ($\dext:S\times
\Re^+_0 \times X \rightarrow S$) is invoked. 
The  new state value depends not only on the input event value but also on the previous state value and the elapsed time since the last transition. 
The new state is calculated as $s_4=\dext(s_3,e,x_1)$ (note that
$ta(s_3) \ge e$). 
No output event is produced during an
external transition.

DEVS models can be coupled modularly \cite{Zeigler2018TheoryFoundations}. A DEVS
coupled model $CN$ is defined by the structure:
\begin{equation*}
CN= \:<X_{self},Y_{self},D,\{M_i\},\{I_i\},\{Z_{i,j}\}, Select>
\end{equation*} 
where:
\begin{description}[noitemsep]
    \item $X_{self}$ and $Y_{self}$ are the sets of input and output values of the
  coupled model.
\item $D$ is the set of component references, so that for each $d\in
  D$, $M_d$ is a DEVS model.
\item For each $d\in D\cup \{self\}$, $I_i\subset (D\cup \{self\})-\{d\}$
is the set of Influencer models on subsystem $d$.
\item For each $i\in I_i$, $Z_{i,j}$ is the translation function,
  where
  \begin{equation*}
      Z_{i,j}:
      \begin{cases}
          X_N\rightarrow X_j &\text{if }i={self}\\
          Y_i\rightarrow Y_{self} &\text{if }j={self}\\
          Y_i\rightarrow X_j &\text{otherwise}
      \end{cases}
  \end{equation*}
\item $Select:2^D\rightarrow D$ is a tie--breaking function for
    simultaneous events. It must verify $Select(E)\in E$, being
    $E\subset 2^D$ the set of components producing the simultaneity of events. 
\end{description}
DEVS models are closed under coupling, i.e., the coupling of DEVS
models defines an equivalent atomic DEVS model \cite{Zeigler2018TheoryFoundations}.  

\section{Extending DEVS for capturing emergent behavior: Design goals and principles}
\label{sec:extending}

So far we have discussed why emergence is an important topic in complex systems. Yet, integrating emergence into an M\&S formal framework presents some nuances. For instance, emergent properties should not be explicitly encoded into a model, for then there would not be \textit{true} emergence at all, no `surprise' factor whatsoever. There are also decisions to be made about what DEVS flavor to take as a departure point for the new formalism.

Classic DEVS has shown sound capabilities to represent a plethora of formalisms~\cite{vangheluweDEVSMUlti}, both in the discrete and continuous realms, and also in the deterministic and stochastic domains. Since its initial introduction~\cite{zeigler1976theory} several new practical needs showed up arising from challenges in the practice of DEVS-based M\&S. These needs motivated the proposal of powerful extensions such as parallel semantics (P-DEVS~\cite{Chow1994ParallelFormalism}), variable structure (DS-DEVS~\cite{Barros1997,Uhrmacher2001}), multi-level dynamics (ML-DEVS~\cite{Steiniger2016}) just to name a few, and also combinations of them.

Yet, we purposely decide to step back from well-known advanced DEVS extensions and frame our approach to Classic DEVS, aiming at answering the following basic question: What could be a minimal formal structure that allows for reasoning about emergence in a DEVS-like model, that is naturally compatible with the original classic formalism?

Thus, in order to organize the formulation of such new formalism we propose a list of \textbf{design goals}:

\begin{enumerate}[noitemsep]
    \item Extend Classic DEVS to allow for dealing with emergence, relying on  multi-level feedback loops between macro and micro-states. \label{a}
    \item Minimize the number of interventions required into Classic DEVS. \label{b}
    \item Preserve the state information hiding property in DEVS models, relying on communication channels to control information sharing between levels.  \label{c}
    \item Offer the usage of micro-macro dynamics as an optional capability, to be either adopted or ignored by the modeler in a flexible way. \label{d}
    \item Provide model heterogeneity, allowing for models with and without emergence to coexist and interact seamlessly. \label{e}
    \item Minimize the impact in terms of changes to the Classic DEVS abstract simulator. \label{f}
    \item Minimize the complexity for updating macro-level states derived from the micro-level models. \label{g}
    \item Foster an intuitive DEVS-style modeling of top-down and bottom-up dynamics. \label{h}
    \item Discourage the explicit modeling of macro-states detached from (or unrelated to) dynamics driven by micro-states. \label{i}
\end{enumerate}

The list above is driven by an underlying purpose to privilege compactness in the formalism, perhaps sacrificing some practical conveniences (e.g.\ parallelism or dynamic structure) which will be the subject of future research. Yet, as it will become clear in \autoref{sec:ebdevs}, we will resort to a flavor of parallelism in the upward communication channel (a collection of \textit{parallel} bottom-up messages in a \textit{bag}).

We shall then introduce a macro-level state in the formalism that is key to identify, analyze, and model emergence. Quoting John Holland ``\textit{aggregates at one level become `building blocks' for emergent properties at a higher level}'' \cite{Holland2014a}(pp. 4-6).

\begin{wrapfigure}{r}{0.5\textwidth} 
  \vspace{-10pt}
  \begin{center}
    \includegraphics[width=0.48\textwidth]{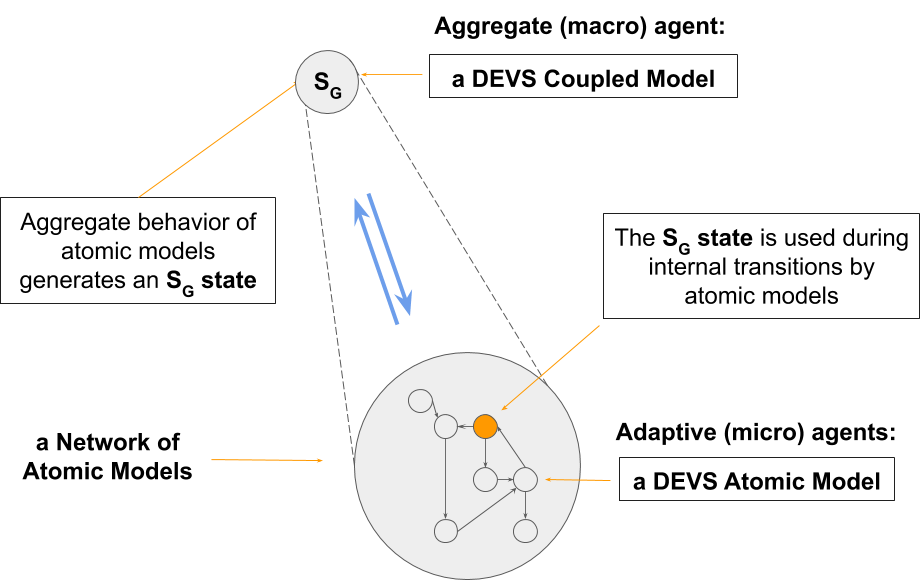}
    \caption{Conceptual approach to equip DEVS with information channels across model levels in the parent-children hierarchy.}
    \label{fig:intuition}
  \end{center}
  \vspace{-10pt}
\end{wrapfigure}

In \autoref{fig:intuition} a conceptual scheme shows the information flow between the macro and micro system levels. 

A (global) \textit{macro-state} $s_G$ appears stored at the coupled model, as an aggregate state that emerges from a network of (individual) \textit{micro-states} at the atomic models (i.e., the building blocks that make emergence possible).

The macro-state shall be produced by a new \textit{global function} applied across micro-states. In fact, in order to preserve information hiding (\autoref{c}) the global function should act only upon what is exposed to it, stemming from the micro-states. We can consider that at any instant in a simulated timeline, atomic models present their environment with a set of visible properties by means of an upward causation mechanism. This should conform a sequence of consistent \textit{collective snapshots} of microscopic states. As a consequence, for any given instant in a simulation there will be a single, unambiguous macro-level state resulting from the latest evaluation of a global function (applied over the latest collective snapshot).

In turn, the macro-level state should be always exposed back to all children models by means of a downward information mechanism, thereby (possibly) influencing micro-level state transitions. Namely, during each atomic model's internal or external transition the macro-level state is made readable to the transition functions to produce the state change. The latter might trigger a subsequent update of the coupled model's aggregate state, via the upward causation mechanism explained above, hence evidencing the desired micro-macro closed loop capability (\autoref{a}). Obviously, a well-defined orchestration mechanism must be defined in order to grant that race conditions can never be created as the result of the multi-level information exchange mechanisms.

Yet, the closed loop mechanism shall not be mandatory (\autoref{d}). It is not mandatory to resort to micro-macro dynamics. It is possible to use the formalism the same way as in Classic DEVS, when multi-level interaction is not required. Namely, the information exchange through upward and downward channels must be optional, and the execution of all transition functions (at both, coupled and atomic models) will treat multi-level information only as an optional piece of data, thus achieving \autoref{e}.

Also, it is worth noting that micro-macro dynamics should be allowed to span across all levels in a model hierarchy. In other words, the capability of pushing messages upwards by an atomic model to its parent must be recursive, hence the coupled model being also capable of pushing messages to its own parent model (i.e., the grandparent of the original atomic one). This can create a cascade of upward causation transitions, spreading information from the bottom-up.

As a consequence, due to the micro-macro loops explained above, and considering the tree-like hierarchical structure of DEVS, every coupled model in a system will have the chance to be influenced (although indirectly) by any atomic model in the system, and vice versa.

  Taking into consideration the \textbf{design goals} explained above, and taking  Classic DEVS as a departure point, we define the following \textbf{design strategy} (refer to \autoref{fig:commodel} to see mathematical elements mapped to a model architecture):

\begin{itemize}[noitemsep]
        \item Equip the coupled model with new elements (functions and sets) at a ``global'' level.

        This is a rather evident design choice, aiming mainly at the goal in \autoref{a}. We will use sub index G for the new elements of the coupled model. In this sense the concept of ``macro'' and ``global'' can be used interchangeably.
       
        \item Add a global state $s_G$ calculated by a global transition function $\delta_G$ triggered exclusively by the arrival of upward messages.
       
        \item Avoid implementing a time advance function at the global level.

        This decision prevents entailing a coupled model with autonomous global behavior (i.e., not driven by micro dynamics), and also prevents creating a type of coupled model that is conceptually very close to what an atomic model is. With this strategy we mainly support goal in \autoref{i} impacting also goals in \autoref{b}, \autoref{e} and \autoref{f}.
       
        \item Allow all new functions and sets to remain undefined, yet producing well-formed models. 

        In the limit case, ignoring all information related to micro-macro dynamics should render a valid Classic DEVS.
       
        \item Adopt a \textit{many-to-one causation} mechanism for the upward micro-macro communication channel. 

          Possibly many atomic models bubble up selected information from their state set during a given simulation cycle (same timestamp). Said information is sent up in the form of $y_{up}$ messages, and accumulated in a bag (or mailbox) $x^b$ at the global level. Upon reception of all $y_{up}$ messages in a simulation cycle, $\delta_G$ is invoked. 
       
        \item Use internal and external transition functions to produce the upward $y_{up}$ messages. 
        State transition functions $\delta_{int}$ or $\delta_{ext}$ calculate both the new  state $s$ \textit{and} the new $y_{up}$ message. We avoid including a new type of $\lambda_{up}$ \textit{upward} output function (i.e.\ imitating the classic $y=\lambda(s)$ scheme) as it is not needed. The reason for this is that the simulation cycle for upward messaging is not decoupled from the cycle of a state transition (as is the case with Classic DEVS messaging).

        \item Adopt a \textit{one-to-many value coupling} mechanism for the downward macro-micro communication channel.

          Every atomic model has read-only access to the global $s_G$ state through a local $s_{macro}$ state (transformed by $v_{down}$), while undergoing the $\delta_{int}$ or $\delta_{ext}$ state transition functions. 
        The triggering of these functions remains the same as in Classic DEVS, now with additional macro-level information made available to decide on micro-level behavior. 
       
        \item Make upward messaging and downward value coupling recursive mechanisms throughout all hierarchical levels.

        Thus, the $y_{up}$ messages and $s_{macro}$ mirrored states (valid at atomic models) have their global counterparts (valid at coupled models), namely  $y_{G_{up}}$ (calculated by $\delta_G$) and $s_{G_{macro}}$ (mirrored from the parent's parent).
       
        \item Make the global state $s_G$ depend on three levels: macro, global, and micro.

        At a coupled model, the global state transition function $\delta_G$ depends on the global state $s_G$, its elapsed time $e_G$, its children's micro-level information (stored at $X^b$) and its parent's macro-level state (mirrored at $S_{G_{macro}}$).

\end{itemize} 

In the following section we introduce the EB-DEVS formalism, derived from the goals and strategies described in this section.

\section{The Emergent Behavior DEVS (EB-DEVS) formalism}~\\
\label{sec:ebdevs}

\begin{figure}[!htb]
\begin{center}
    \vspace{-25pt} 
    \includegraphics[width=\textwidth]{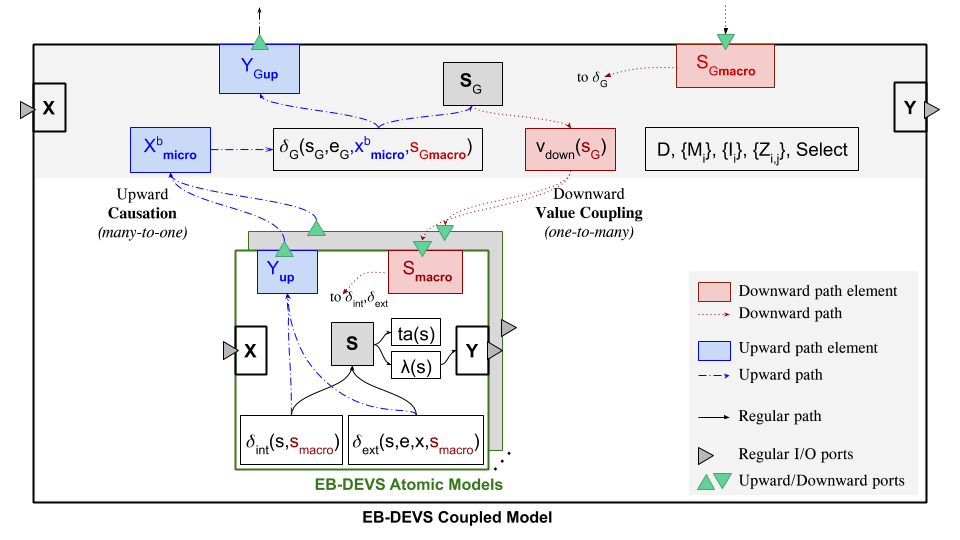}
    \caption{Architecture and communication model for EB-DEVS.} 
    \label{fig:commodel}
  \end{center}
  \end{figure}

The \fullref{fig:commodel} shows how atomic and coupled models interact with each other.
Upward causation events are written in the output ports $Y_{up}$. They are read by the coupled model in the $X_{micro}^b$ input bag port.

The upward causation output port in the coupled model is called $Y_{G_{up}}$. Conversely, it receives information from the higher-level models through the $S_{G_{macro}}$ input port via value couplings. These input and output ports relationships is a common denominator between levels giving consistency through the model hierarchy.

\subsection{EB-DEVS Atomic Model formal definition}  
\label{ssec:formal}


Several differences compared with the atomic DEVS model exist, some of them have been discussed in \autoref{sec:extending}, regarding the new communication channels. These changes enable the top-down and bottom-up information sharing process. But the new available information  needs to be taken into account by the state transition functions. This is one of the major changes in the atomic model. State transition functions $\dint$ and $\dext$ use the parent's model state as a parameter for the state change. Their domain and co-domain have changed accordingly, and their outputs are used to communicate the results to the parent model.

Let us now look at the Atomic EB-DEVS model definition.  
\begin{align*}
  M = &<\overbrace{X,
    Y,
    S,
    ta,
    \dint,
    \dext,
    \lambda,}^{\text{Classic DEVS}}~\overbrace{Y_{up},
    S_{macro}>}^{\text{EB-DEVS}}
\end{align*}

Where,

\begin{description}[noitemsep,labelsep=1em, align=left, labelwidth=1in,labelindent=1cm]

    \item[$X$] is the set of input events.
    \item[$Y$] is the set of output events.
    \item[$S$] is the set of states.
    
    
    \item[$ta: S \rightarrow \Re_{0}^{+} \cup \infty$] is the time advance function that determines the lifespan of a state.

    \item[$\dint:S \times S_{macro} \rightarrow S \times Y_{up}$] is
        the internal transition function which defines how a state of the model changes autonomously (when the elapsed time reaches the lifetime of the state). The second output value in the tuple, defined
        in $Y_{up}$ is the information for the parent to compute $\delta_G$ (defined in the next section).
    
    \item[$\dext: S \times \Re_{0}^{+} \times X \times S_{macro} \rightarrow S \times
        Y_{up}$] is the external transition function which defines how an input
        event changes the current state of the model. The value for the output
        port $Y_{up}$ is defined as in the internal transition.

    \item[$\lambda: S  \rightarrow Y$] is the output function.

    \item[$Y_{up}$] is the set of output events directed to the parent model.
    
    \item[$S_{macro}$] is the (value coupled) set of parent's states.
    
\end{description}

\subsection{The EB-DEVS Coupled model Formal Definition}~\\
\label{sec:coupleddef}

Classic DEVS coupled models are containers that enable hierarchical compositions of other models. However, they have no behavior nor state of their own.
In EB-DEVS, an upward communication channel is defined to convey messages coming from the set $Y_{up}$ (at the micro level) towards the set $X_{micro}^b$ (at the macro level) enabling an upward causation mechanism. It allows for the calculation of a  global state in $S_G$ that depends on local information recollected at $X_{micro}^b$, and is calculated by the global transition function $\delta_G$.

Also, as an argument of $\delta_G$ is $S_{G_{macro}}$ which allows for accessing the coupled model's parent global state (downward information channel from upper layers).

To compute the new macro-level state $s_G \in S_G$, the function $\delta_G$ also uses the current value of $s_G$ and its elapsed time $e_G$ (similarly to what a Classic DEVS external transition function does).

Each calculation of $\delta_G$ can also produce upward output messages through the $Y_{G_{up}}$ set.

For the downward communication channel (from parent to children) we define a value coupling mechanism by means of the 
$v_{down}$ function. It has access to the global state in $S_G$ and, when invoked by one or more children, it can restrict the amount of global information made visible to the lower layer.

\begin{align*}
  {CN} & = \overbrace{<X_{self}, 
    Y_{self},D, 
    \{M_i\},
    \{I_i\},
    \{Z_{i, j}\},
  \text{Select} ,}^{\text{Classic DEVS}}  \overbrace{X^b_{micro}, 
                   Y_{G_{up}},
                   S_{G_{macro}},
                   S_G, 
                   v_{down},
                 \delta_G>}^{\text{EB-DEVS extension}} 
               \end{align*}

The new coupled model preserves the original elements of Classic DEVS coupled models, and adds six new elements.

\begin{description}[noitemsep,labelsep=1em, align=left, labelwidth=1in,labelindent=1cm]

  \item[$X_{self}$] is the set of input values.

  \item[$Y_{self}$] is the set of output values.
  \item[\rm D] is the set of component references.
  \item[$\{M_i\}$] for each $d\in D$, $M_d$ is a DEVS model.

  \item[$\{I_i\}$] is the set of influencer models for each subsystem.

  \item[$\{Z_{i, j}\}$] for each $i\in I_d$, $Z_{i,d}$ is the translation function.

  \item[\rm Select] is the tie breaking function for simultaneous events.

  \item[$X^b_{micro}$] is a mailbox input port for the information events sent by the atomic models.
  \item[$Y_{Gup}$] is an output port for the information events sent towards the parent.
  \item[$S_G$] is the set of states that the coupled model can adopt.
  \item[$S_{Gmacro}$] is the parent's set of states. It can be $\oslash$ in case of not having a parent.
  \item[$v_{down}: S_G \rightarrow S_{macro}$] is the downward value coupling function that provides the global information to its children.
  \item[$\delta_G: S_G \times \Re_{0}^{+} \times X^b_{micro} \times S_{Gmacro} \rightarrow S_G \times
    Y_{Gup} $] is a function that computes a new macro state $S_G$ based on its own state, the elapsed time for its last state change, the messages $X^b_{micro}$ arrived from its micro components and its parent's macro state $S_{Gmacro}$. It also computes the upward-causation event (a value in the $Y_{G_{up}}$ set) towards its parent. The cascade of upward causation events can eventually climb up in the hierarchy, possibly (but not necessarily) until the root coupled model.
\end{description}

\subsection{Theoretical Properties of EB-DEVS}
\label{ssec:prop}

In this section we show three core theoretical properties of EB-DEVS\@. The closure under coupling property enables us to define a single, universal EB-DEVS abstract simulator for any modular/hierarchical composition of atomic and coupled EB-DEVS models.
Also, a definition for the notion of legitimacy of EB-DEVS must be provided, to delimit the class of EB-DEVS models that can produce correct simulations (in the sense of guaranteeing a physically realizable representation of time). We will approach this definition by relying on a third property, the bisimulation between EB-DEVS and DEVS\@. 

\begin{wrapfigure}[18]{r}{0.6\textwidth} 
  \vspace{-30pt}
\begin{center}
    \includegraphics[width=0.58\textwidth]{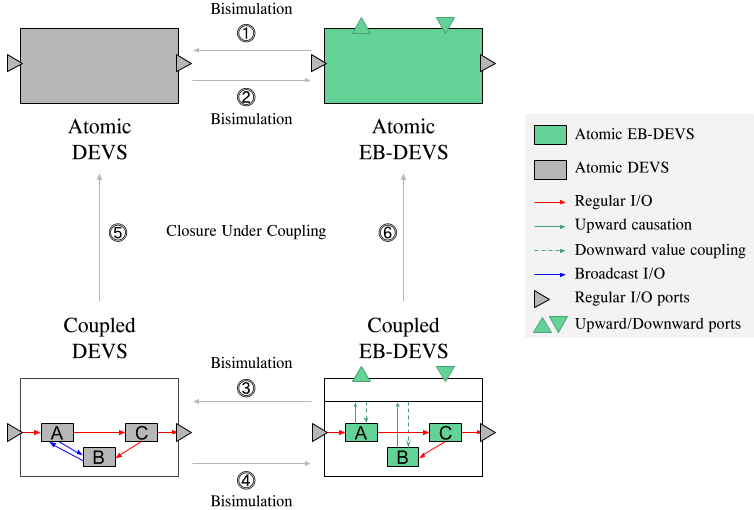}
    \caption{Formal relations and possible transformations between DEVS and EB-DEVS models.}
    \label{fig:model_transf}
\end{center}
\vspace{15pt}
\end{wrapfigure}

\autoref{fig:model_transf} synthesizes the relations between the closure under coupling and bisimulation properties between DEVS and EB-DEVS models, both atomic and coupled.

The numbered arrows depict source-destination pairs. For instance, arrow 4 represents the fact that it is always possible to find a Coupled EB-DEVS model that bisimulates a given Coupled DEVS model, and arrow 6 represents that given a Coupled EB-DEVS model we can always find an equivalent Atomic EB-DEVS model. We refer the reader to the appendices for the formal proofs of closure under coupling (\autoref{sec:cuc}), bisimulation (\fullref{sec:bisim}) and legitimacy (\fullref{sec:legit}).

\subsubsection{Abstract Simulator for EB-DEVS}~\\

The DEVS simulator (also, the DEVS \textit{abstract} simulator) was first introduced, and later evolved by Bernard Zeigler  \cite{zeigler1976theory,zeigler2000theory,Zeigler2018TheoryFoundations}.
It allows generating unambiguous behavior (state and output trajectories) for any given system specification and initial conditions. Additionally, the formalism enforces rules to make the system specification and its resulting behavior match with each other. The DEVS  abstract simulator defines how models are executed. For instance, it ensures that the output function is invoked before the internal transition, that the select function is applied in case of a need for tie-breaking, and how to dispatch the output messages between models. In this section we extend the DEVS abstract simulator to express the new kind of behavior required for EB-DEVS to work. 

The corresponding EB-DEVS abstract simulator inherits from DEVS the tree-like structure and the message passing mechanism. As in DEVS, there are three types of processors: the \textit{EB-DEVS-root-coordinator} (which has no model associated with it and forms the root processor of the tree), the \textit{EB-DEVS-coordinators} (which are associated with coupled models, and represent the inner nodes of the tree) and the \textit{EB-DEVS-simulators} (which are associated with atomic models and represent the leaves of the tree). The EB-DEVS-root-coordinator initiates and controls the simulation cycles, it sends the initialization messages and forwards top-level messages to the corresponding lower-level processor. Correspondingly,  processors ( EB-DEVS-coordinators and EB-DEVS-simulators) communicate information between model levels (top-down, bottom-up, or the same level). Communication is done with messages to signal initialization, external transitions, internal transitions, output events, or when a model has ended executing its state transition (i.e.\ *-messages, x-messages, y-messages, and done-messages).

The main modifications brought into the abstract simulator are  the exposure of macro-level states to the micro-level atomic models, and the execution of the coupled model's transition function $\delta_G$.

For a detailed explanation about how this is implemented, please refer to the \autoref{ssec:abstract}, where we show how the EB-DEVS abstract simulator's  processors are extended.

\section{Case studies}
\label{sec:cases}

In this section we adopt EB-DEVS to model three types of systems with micro-macro level interactions. The purpose is to explore both the potential and limitations of the formalism.

First we will discuss the implementation of the classic epidemiological Susceptible Infected Recovered (SIR) model \cite{Kermack1927AEpidemics}. We will show how EB-DEVS can implement Classic DEVS models (without multi-level interactions) and how easy it is to extend its behavior to take advantage of multi-level closed-loop dynamics.

The second model is the classic Reynold's Boids model \cite{Reynolds1987} and two additional extensions. These models will showcase how it is possible to represent complex behavior by integrating the multiple levels of the system. We will resort to EB-DEVS to exploit indirect communication, adaptive behavior, and spontaneous organization as model features.

Finally, we implemented a recent mitochondria model, based on Dalmasso's agent based model \cite{Dalmasso2017Agent-basedHeterogeneity}. In this case, we test the limits of the formalism by modeling a system that calls for variable model structure. We will show how it can be approached with EB-DEVS, what additional features would be required and how indirect communication can be of help to circumvent the variable structure issue.

\subsection{A SIR Model}

The SIR model is a classical epidemiological compartmental model proposed by Kermack and McKendrick and further extended by Hoppensteadt \cite{Kermack1927AEpidemics,Hoppensteadt1974AnModel} and many others.
  
Susceptible, Infected, and Recovered compartments in a given population evolve over time influencing each other. The model presents strong assumptions. All Susceptible (S) become Infected (I) at a rate $\beta I$, while all Infected become Recovered (R) at a rate $\gamma I$. The parameter $\beta$ represents the average number of contacts per person multiplied by the probability of transmission at the moment of each contact. On the other hand, $\gamma$ describes the rate of recovery.


This simple model can be used to study how a contagious  disease spreads in terms of the evolution of the amount of infected population, duration of the epidemic, effect of social distancing policies, impact of vaccination campaigns, etc. 

The SIR model has been extended and studied extensively by changing the types of compartments \cite{Allen1994}, the connectivity network \cite{ferreyra2019sir} (by using mean-field theory), or the conditions of  the infection process.

Investigation through Ordinary Differential Equations (ODEs) is one of the possible macroscopic approaches to study this model. 
The following is a representation in the form of a set of ODEs parametrized with the above mentioned rates:

\begin{equation}
\label{eqn:SIR}
  \begin{array}{rl}
    \dfrac{dS(t)}{dt} & = -S(t)\dfrac{\beta I(t)}{N} \\[6pt]
    \dfrac{dI(t)}{dt} & = S(t)\dfrac {\beta I(t)}{N}-\gamma I(t) \\[6pt]
    \dfrac{dR(t)}{dt} & = \gamma I(t) \\ [6pt]
    N & = S(t)+I(t)+R(t)
  \end{array}
\end{equation}

Agent-based discrete event models are also an effective approach when microscopic phenomena needs to be taken into account. Interactions among agents can be modeled as a graph where pairs of nodes are connected if there is a potential of interaction. One option is to focus on the degrees of each node, that is, the number of encounters each person may have with others. A common approach is to model such social interactions with a Configuration Model (CM) type of network, widely used in social science research. A CM allows building complex networks with an arbitrary sequence of node's degrees (for an in-depth explanation of Configuration Models see the works of Barabási \cite{barabasi2016network} and Bollobas \cite{bollobas2001random}). In this section we will call SIR-CM the SIR model that spans over a Configuration Model graph.

\subsubsection{Modeling a SIR-CM with EB-DEVS}~\\
\label{ssec:sir_map}
We present a possible implementation of the SIR-CM model with EB-DEVS\@. As we will see, the most interesting part of this example comes when some emergent property is detected at the system level, such that it becomes reflected at the agents' level.

We model the environment as a coupled model. It contains atomic models (agents) connected by means of a network \textit{N} built as a CM graph. The degree sequence of the CM follows a Gamma distribution with parameters ${k}=10, \: \theta=1$.

Each atomic model has a state variable denoting the S, I or R stages of the agent. Each pair of interacting atomic models are bidirectionally connected through input and output ports according to the topology of \textit{N}. Atomic models have a single input port to receive messages coming from (possibly many) other models. The rules that define the dynamics of each agent are:
\begin{itemize}[noitemsep]
    \item A Susceptible agent changes its state to Infected when it receives a message through its input port (signaling an infection).
    \item An Infected agent sends infection messages through its output ports (one at a time, and never more than one at the same time). It may infect more than one agent during its infectious period (before recovering).

    \item An Infected agent will recover from the disease changing its state to Recovered, and never get infected again (even if it receives further infection messages from other agents). This models the immunization of the agent.
\end{itemize}

The rates at which agents get infected and recover are in accordance with the SIR structure in the form of ODEs in  \autoref{eqn:SIR}.
To translate those rates into the agent's state transition functions, we define two exponential clocks that will trigger individual infections and recoveries. An agent will stay in the Infected state for a period sampled from an exponential distribution $exp(\lambda = 1/\gamma)$. After that, it will transition to  Recovered. The same agent will infect other agents by following a clock with exponential distribution $exp(\lambda = 1/(\#neighbors \cdot \beta))$.


The internal transition determines if an atomic model can either infect another model or recover, changing its state depending on the clock's events order. The situation of having $i \in (1, \dots, n)$ concurrent exponential clocks with parameter $\lambda_i$ is known as an \textit{exponential
race}. To solve this problem we define $\xi$ as the minimum in a set of $\xi_i$ exponentially distributed variables of parameter $\lambda_i$. The parameter for the exponential distribution of $\xi$ is $\lambda = \sum_i^n \lambda_i$.
The probability that $\xi_i$ be the triggered event among $\xi$ has probability $P(\xi = \xi_i) = \lambda_i/\lambda$ 
(see Bocharov's \textit{Queuing Systems} \cite{Bocharov2011QueueingTheory} book, lemmas 1.2.2 and 1.2.3, for the proof).




We define the EB-DEVS primitives as follows.
\begin{itemize}[noitemsep]
    
  \item The time advance function $ta$ for a Susceptible or a Recovered agent returns $\infty$, while for an Infected agent it returns values sampled from the exponential distribution \[
  exp(\lambda=\frac{1}{\#neighbors \cdot \beta} + \frac{1}{\gamma})
  \]
  
  \item The external transition function $\dext$ changes the agent's state upon reception of an infection message. If the agent is Susceptible, it changes its state to Infected.

     

  \item The internal transition function $\dint$ acts only on the Infected state.
  It changes the state from Infected to Recovered with probability
     \[
     P=\frac{\#neighbors \cdot \beta}{\#neighbors \cdot \beta + \gamma}
     \]
    (remaining in Recovered until the end of the simulation ignoring any other event).
    Conversely, it will remain in state Infected with probability $1-P$.

    \item The output function $\lambda$ will emit an infect message through a randomly selected output port (uniform distribution) only if the agent is in Infected state.

    \item The relationship between the $\dint$ and $\lambda$ functions can be stated as follows. Consider that $s_{i+1}=\dint(s_i)$ after $ta(s_i)$ units of time. There are two possible cases that are selected randomly at the $\dint$ function. In one case, at the \textit{next internal transition} the infected agent shall recover according to $s_{i+1}=\dint(s_i) = R$ and $ta(R)=\infty$ (no further output message will be issued). Otherwise, at the \textit{next internal transition} the agent shall infect a neighbor and remain infected, according to $\lambda(I)=\text{infect}$ and $s_{i+1}=\dint(s_i)=s_i=I$.
    
    \item Each state change is informed by every agent to the coupled model with the $y_{up}$ output port \textit{(many-to-one upward causation)}.
    
    \item The coupled model aggregates the states of its dependants, by means of its $\delta_G$ function, calculating the number of agents for each compartment (S, I and R). This information is \textit{not} made available to the atomic models (i.e., the \textit{one-to-many downward value coupling} would return a null value if invoked by an agent).
    
\end{itemize}

\subsubsection{Formal instance of the SIR-CM model}~\\
We give the formal definition of the SIR-CM model using EB-DEVS. We define the following atomic model:
$$ Agent = \: <X, Y, S_A, ta, \dint, \dext, \lambda, Y_{up} , S_{macro}> $$

\textbf{Note: } we name the agent`s state set $S_A$ instead of the usual $S$ to avoid a naming clash with the Susceptible compartment.

\begin{multicols}{2}
\begin{description}[noitemsep]
    \item $X = Y = \{\text{Infect}\}$
    \item     $S_A = \{S, I, R\}$
    \item     $ta(S) = ta(R) = +\infty$
    \item     $ta(I) = exp(\lambda=\frac{1}{\#neighbors \: \beta} + \frac{1}{\gamma})$
    \item     $\lambda(I) = \text{Infect}$
    \item     $\lambda(R) = \lambda(S) = \oslash$
    \item     $\dint(S, \cdot) = S$
    \item     $\dint(I, \cdot) = \text{if } {unif}(0, 1) < \frac{\#neighbors \: \beta} {\#neighbors \: \beta + \gamma} $ 
    
\hspace{1cm} then $ R \text{ else } I$

    \item     $\dint(R, \cdot) = R$
    \item     $\dext(S, e, \text{Infect}, \cdot) = \dext(I, e, \text{Infect}, \cdot) = I$
    \item     $\dext(R, e, \text{Infect}, \cdot) = R$
    \item     $Y_{up} = \{S, I, R\}$
    \item     $S_{macro} = \oslash$ 
\end{description}
\end{multicols}

We define the \textit{`Environment'} coupled model, as a container for the atomic models (or \textit{`agents'}). The network \textit{N} (see \autoref{ssec:sir_map}) defines the values for $I_i$ and $Z_{i,j}$.

\begin{equation}
\begin{array}{l}
    {Environment} =  <X_{self}, Y_{self}, D, \{M_i\}, \{I_i\}, \{Z_{i, j}\},\\
      \quad \text{Select}, X^b_{micro}, Y_{G_{up}}, S_{G_{macro}}, S_G, v_{down}, \delta_G>
\end{array}
\end{equation}

\begin{multicols}{2}

\begin{description}[noitemsep]
    \item $X_{self}= Y_{self}=\oslash$
    \item $D=\{1,\,\dots\, , N\}$
    \item $\{M_i\} = \{\text{Agent}_i\}$
    \item $\{I_i\} = \text{CM-graph-connectivity}$
    \item $\{Z_{i, j}\} = \{ {Infect } \} \rightarrow \{ {Infect} \}$ 
    \item $\text{Select} = \text{sort by } i \in D$
    \item $X^b_{micro} =\{S, I, R\} $
    \item $Y_{G_{up}} = S_{G_{macro}} = \oslash$
    \item $S_G = \{(x, y, z)  | x, y, z \in \mathbb{N} \}$
    \item $v_{down} = \oslash$
    \item $\delta_G(s_G, e, s, s_{G_{macro}}) = (\text{\# of agents with state S}+1, \\
        \text{\# of agents with state I} , \\
        \text{\# of agents with state R}) $ 
    \item $\delta_G(s_G, e, I, s_{G_{macro}}) = (\text{\# of agents with state S}, \\
        \text{\# of agents with state I} +1, \\
        \text{\# of agents with state R}) $
    \item $\delta_G(s_G, e, R, s_{G_{macro}}) = (\text{\# of agents with state S}, \\
        \text{\# of agents with state I} , \\
        \text{\# of agents with state R}+1) $

\end{description}
\end{multicols}

\subsubsection{SIR-CM-V: Extending the SIR-CM model with vaccination} ~\\

We extend the SIR-CM model to provide atomic models with aggregated information of the rate of infection. This information is later used by the atomic models to decide whether to vaccinate or not.

A vaccination campaign is deployed to prevent infection of a Susceptible agent. To achieve this, we use an \textit{outbreak} emergent property.

In this case the primitives are implemented as follows:
\begin{itemize}[noitemsep]
    \item The internal transition, time advance, and output functions remain the same as in the SIR-CM model.
    \item The external transition function $\dext$ is extended to read the emergent outbreak information made available by the coupled model. If the outbreak variable crosses a given threshold, when the agent detects this property then decides to avoid current and future infections by setting its \textit{vaccinated} flag. 
    
    \item The $\delta_{G}$ function computes $s_G$ as the growth rate for the Infected compartment, discretizing time into regular time bins. As we have the number of infected agents in the infected compartment for time bin $N-1$ and $N$ (being $N$ the latest time bin), we can calculate the discrete derivative indicating the rate of growth or decay of infected agents.
    \item The value-couplings function $v_{down}$ shares the rate of growth with the atomic models.
    \item The $y_{up}$ output message shares the new state with the coupled model.
\end{itemize}

\subsubsection{Formal instance of the SIR-CM-V model including vaccination}~\\

Using EB-DEVS we present the SIR-CM-V model. It is based on the SIR-CM and it uses downward information to model a vaccination campaign for preventing infections.

The macro-level state stores the amount of agents at each compartment. It then computes the increase or decrease of infected agents per time unit. This decreases or increases the rate used by the agents to detect if there is an \textit{outbreak} of the infection. Susceptible agents receiving an \textit{infect} message while the \textit{outbreak} emergent property is present will become \textit{"vaccinated"}, staying in Susceptible state until the end of the simulation. The Susceptible compartment is split into the ones that got `vaccinated' ($S_v$) and the ones that have not ($s_{\neg v}$). This is used to avoid future infections after the \textit{outbreak} property is deactivated. We use $S$ as a wildcard in cases where there is no distinction between the two states.

$$ Agent = <X, Y, S_A, ta, \dint, \dext, \lambda, Y_{up} , S_{macro}> $$

\textbf{Note: } we name the agent state set $S_A$, to avoid naming issues with the Susceptible compartment.

\begin{multicols}{2}
  \begin{description}[noitemsep]
    \label{desc:sir-cm-v}
  \item $X = Y = \{\text{Infect}\}$
  \item     $S_A = \{S_{\neg v}, S_v, I, R\}$
  \item     $ta(S) = ta(R) = +\infty$
  \item     $ta(I) = exp(\lambda=\frac{1}{\#neighbors \: \beta} + \frac{1}{\gamma})$
  \item     $\lambda(I) = \text{Infect}$
  \item     $\lambda(R) = \lambda(S) = \oslash$
  \item     $\dint(S, \cdot) = S$
  \item     $\dint(I, \cdot) = \text{if } {unif}(0, 1) < \frac{\#neighbors \: \beta} {\#neighbors \: \beta + \gamma} $ \\ then $ R \text{ else } I$

  \item     $\dint(R, \cdot) = R$
  \item     $\dext(S_{\neg v}, e, \text{Infect}, n) = \text{if} \, n < \text{vaccination threshold} \\ \quad \text{then}
    \: I \: \text{otherwise} \: S_v$
  \item     $\dext(S_{v}, e), \text{Infect}, n) =  S_v$
  \item     $\dext(I, e, \text{Infect}, n) = I$
  \item     $\dext(R, e, \text{Infect}, n) = R$
  \item     $Y_{up} = \{S, I, R\}$
  \item     $S_{macro} = \mathbb{N}$ \\ \\ 
\end{description}
\end{multicols}

We define the coupled model as a container of the atomic models named \textit{'agents'}. The coupled model contains the coupled network details for the connections of the agents with the coupled model and the rest of the model. In this case, we also define the value-couplings and internal / external transition functions to use this information for their transition.

\begin{equation}
\begin{array}{l}
    {Environment} = \: <X_{self}, Y_{self}, D, \{M_i\}, \{I_i\}, \{Z_{i, j}\},\\
      \quad \text{Select}, X^b_{micro}, Y_{G_{up}}, S_{G_{macro}}, S_G, v_{down}, \delta_G>
\end{array}
\end{equation}

\begin{multicols}{2}
  \begin{description}[noitemsep]
    \item $X_{self}= Y_{self}=\oslash$
    \item $D=\{1,\,\dots\, ,N\}$
    \item $\{M_i\} = \{\text{Agent}_i\}$
    \item $\{I_i\} = \text{CM-graph-connectivity}$
    \item $\{Z_{i, j}\} = \{ {Infect } \} \rightarrow \{ {Infect} \}$ 
    \item $\text{Select} = \text{sort by } i \in D$
    \item $X^b_{micro} =\{S, I, R\} $
    \item $Y_{G_{up}} = S_{G_{macro}} = \oslash$
    \item $S_G = \{(x, y, z)  | x, y, z \in \mathbb{N} \}$
    \item $v_{down} = \text{Change (increase or decrease) in the} \\ 
    \text{\# of agents with state I}$
    \item $\delta_G(s_G, e, S, s_{G_{macro}}) = (\text{\# of agents with state S}+1, \\
      \text{\# of agents with state I} , \\
      \text{\# of agents with state R}) $ 
    \item $\delta_G(s_G, e, I, s_{G_{macro}}) = (\text{\# of agents with state S}, \\
      \text{\# of agents with state I} +1, \\
      \text{\# of agents with state R}) $
    \item $\delta_G(s_G, e, R, s_{G_{macro}}) = (\text{\# of agents with state S}, \\
      \text{\# of agents with state I} , \\
      \text{\# of agents with state R}+1) $ \\
  \end{description}
\end{multicols}

\subsubsection{Simulation Experiments}~\\

For the SIR-CM and SIR-CM-V models, 500 atomic models were instantiated with ten percent of the agents starting the simulation with infected state while the remaining with susceptible state. We used a Configuration Model network with degree sequence $\Gamma(\alpha=10, \beta=1)$.

The simulation models were run 50 times each with the same parameters using different  seeds for the random number generation.

\begin{figure}[ht]
\begin{minipage}[t]{.45\textwidth}
\raggedleft
    \centering
    \includegraphics[width=\columnwidth]{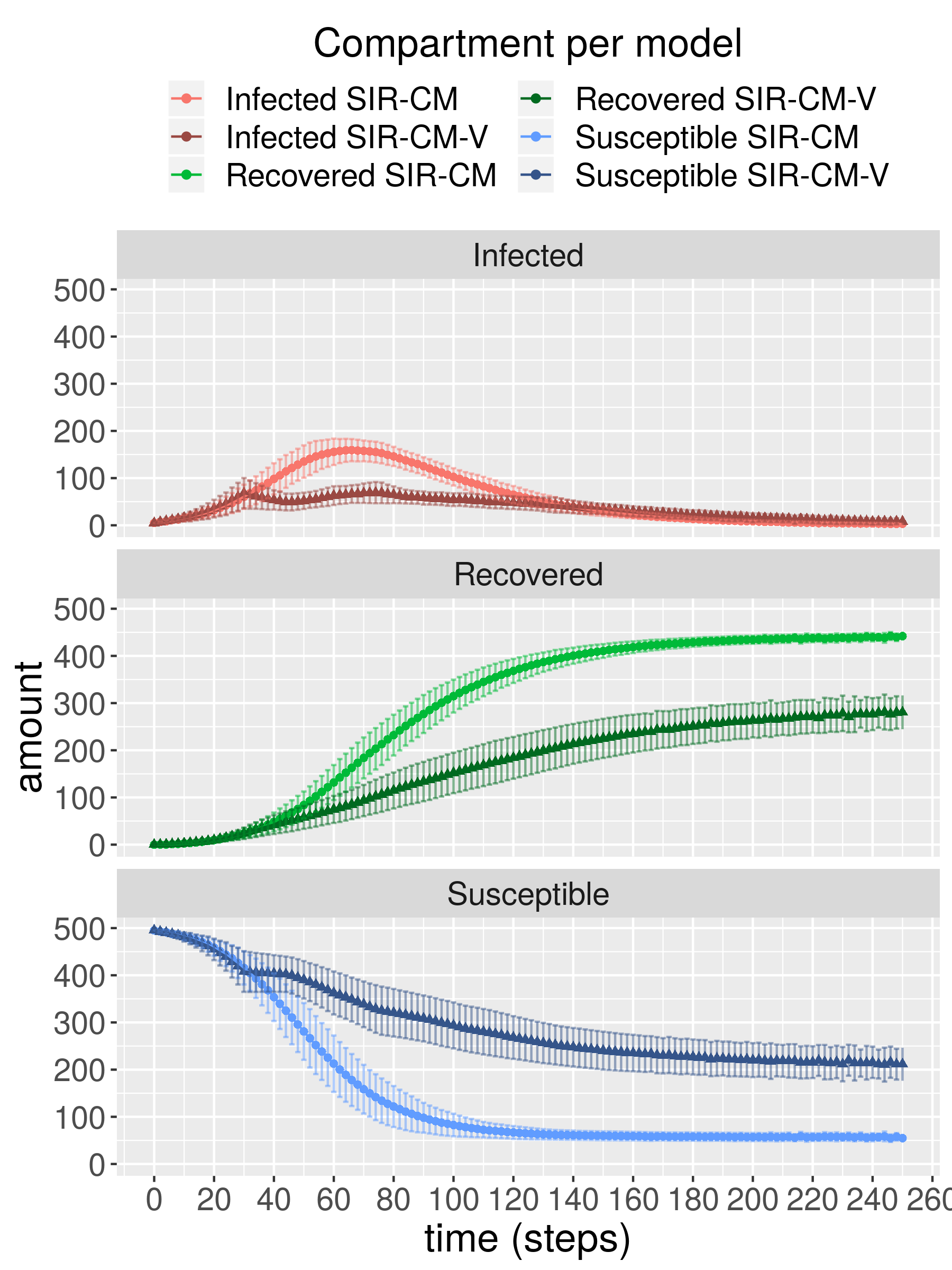}
    \captionof{figure}{SIR-CM model vs. SIR-CM-V model. The x-axis presents time and y-axis the amount of agents in each compartment for each model version. Points present the mean value for the runs while the crossbars present the standard deviation.}
    \label{fig:sir_comparison}
\end{minipage} 
\hfill
\begin{minipage}[t]{.45\textwidth}
\raggedright
    \centering
    \includegraphics[width=\columnwidth]{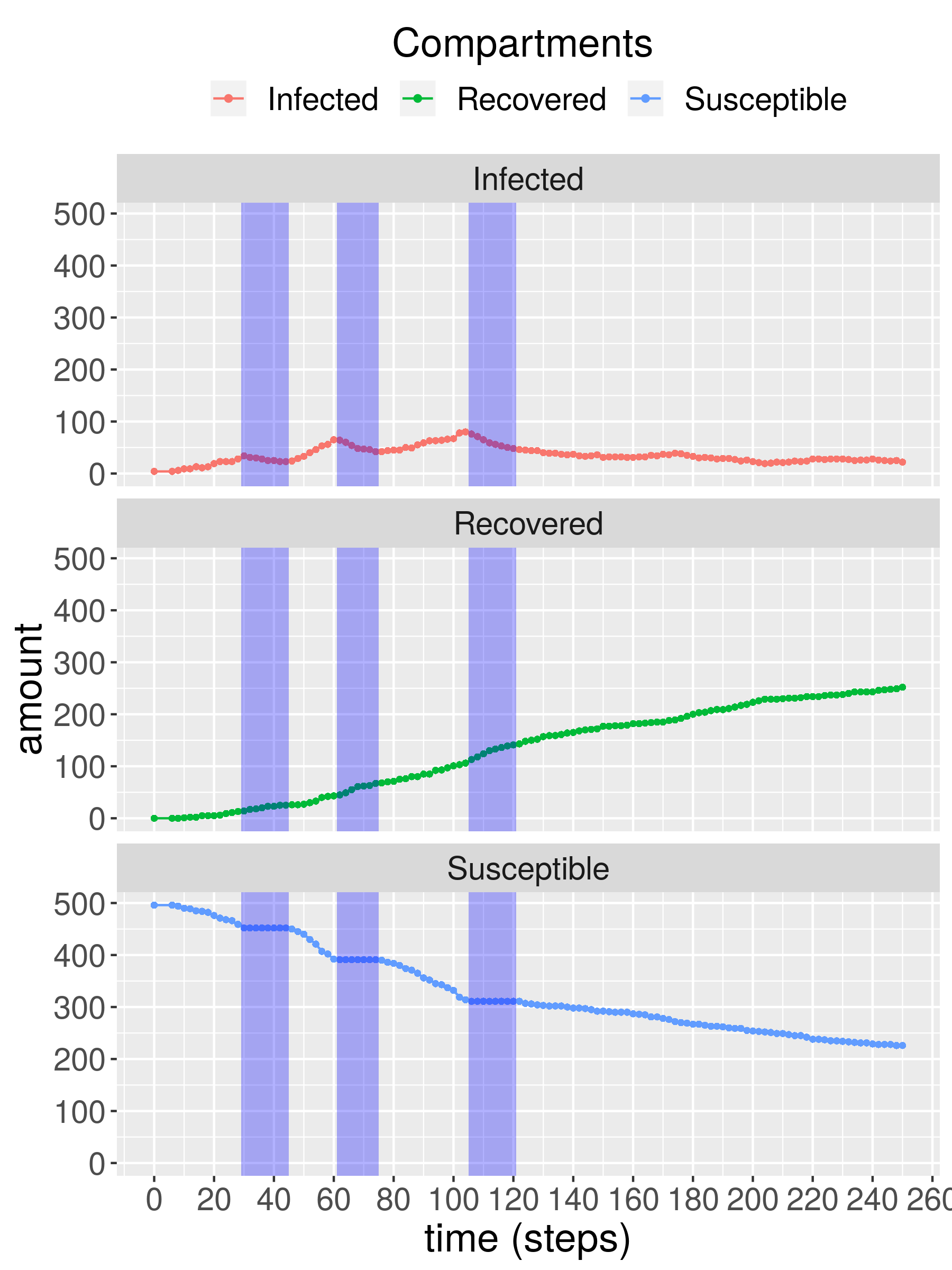}
    \captionof{figure}{SIR-CM-V Model execution. The x-axis shows time. The y-axis presents the amount of agents in each compartment. Vertical lines show the instants at which the vaccination event was active.}
    \label{fig:sir_vacc_single}
\end{minipage}
\end{figure}

In \autoref{fig:sir_comparison} we can see how the two models compare over time. The SIR-CM model presents the expected evolution of a  classic SIR model, the SIR-CM-V model drops the number of infected agents while reaching the outbreak emergent property making it diverge from SIR-CM model. 

We can observe the effect of the vaccination events in \autoref{fig:sir_vacc_single}. Vertical blue boxes highlight the moments where the outbreak emergent property is present. At the same time of detection of the property, susceptible agents ignore the infection message thus preventing their infection.

\subsubsection{Discussion}~\\

We presented a SIR-CM model and its extension, the SIR-CM-V model. Given the model implemented in EB-DEVS, extending it for the use of micro-macro level dynamics proved straightforward. The simplicity of modeling with EB-DEVS shows us how practical it can be to add emergent behavior to existing models.

Yet, extending model behavior to include complex conditions based on a state variable can bring about some modeling issues. As models grow, the conditions needed to support their behavior become more complex, potentially generating too many branches of conditions. Furthermore, if this behavior is bound to macro-level variables, the requirements for their generation can result in  a model re-engineering. During the implementation of the SIR-CM-V model, we incorporated a simple behavior related to an aggregated state variable for the modeling of vaccination in the SIR-CM model. While working with a simple model like SIR-CM-V, if we consider an analogous Classic DEVS implementation it would require rethinking the whole model, recreating its connectivity and probably the atomic models' external transitions. In \autoref{desc:sir-cm-v}, we see how the modifications are restricted to the $\delta_G$ (for the computation of the aggregated macro state), the value couplings (for the  downward communication of the macro state), and the $\dint$ functions (for the actual usage of the newly available information in the state transition).

\subsection{Boids}

The Boids model defined by Reynolds \cite{Reynolds1987} is a well known distributed behavior model that has been extensively used to showcase modeling methodologies and capabilities \cite{Szabo2015,Chan2011,seth2008measuring}. This model offers a good platform to benchmark models that target \textit{emergence} as a key aspect. 

In this section we will study the Boids model and how emergent properties appear in its execution. 
Furthermore, we will explore how indirect communication, adaptive behavior and emergent properties can be exploited to produce self organizing behavior in the birds' movement relative to their flock.

In Reynold's Boids model, birds move in a two dimensional space with constant speed and varying heading according to their position relative to  neighboring birds. Those birds within a given radius are considered neighbors. 

Three rules define how the heading is adapted dynamically:

\begin{description}[noitemsep]
  \item[Separation:] if two neighboring birds are closer than a given distance, move away from the closest bird.
  \item[Alignment:] if there is no need to separate, drive bird's direction towards the average heading of the neighboring birds.
  \item[Cohesion:] finally steer bird's direction towards the center of mass of the neighboring birds.
\end{description}

The environment is a 2-dimensional grid with periodic boundary conditions. 
This means that when a bird hits the border it will re-enter through the other side as it would in a torus ring shaped space.

\subsubsection{Mapping to EB-DEVS}~\\

Our implemented Boids model is a discrete time model with a \textit{Flock} EB-DEVS coupled model that contains \textit{Birds} EB-DEVS atomic models.
     
The Bird's state is defined by their \textit{heading} and a pair of \textit{x,y} coordinates in a continuous 2D space. We assume constant velocity (one spatial unit per time unit) in the desired direction.

The $\dint$ function moves the Bird in space by advancing the position towards the heading direction. In order to determine the nearest neighbor and the flock-mates, the Bird accesses the macro variable stored in $s_{macro}$.
As this model uses only indirect influence among agents, neither the $\dext$ function nor the $\lambda$ output function are implemented.

In its internal transition a Bird will first ask for the closest neighbor to be able to execute the three Reynold's rules. The search for the closest neighbor, the set of closest neighbors, and the proximity graph spanned by the neighboring proximities are resolved by querying a \textit{Radius Neighbors Regressor (RNR)} \cite{Mitchell1997MachineLearning}, a Machine Learning algorithm used to classify points in the euclidean space. We used a RNR modification to support spaces with periodic boundary conditions.

The $\dint$ function will work as specified in the following pseudocode.

\begin{lstlisting}[style=base,linewidth=\columnwidth,breaklines=true,
    basicstyle=\ttfamily\footnotesize,
    xleftmargin=2em,
    numbers=left,
    stepnumber=1,
    tabsize=2,
    firstnumber=1,
    numberstyle=\tiny,
    numberfirstline=false,
    label={lst:intboids},
    caption={Pseudocode of the Boids internal transition}]
function intTransition:
    closest = $v_{down}$[closest]
    if closest is none:
        advance()
        return
    if euclidean.dist between closest.coord and self.coord < Parameters.MAX_DIST:
        separate()
    else:
        align_with($v_{down}$[neighbors])
        cohere_with($v_{down}$[neighbors])
    advance()
    return 
\end{lstlisting}

The separation is done by rotating the heading angle ($\alpha$) away from the closest neighbor. Using $\alpha_{self} - \alpha_{closest}$ for the delta increment of the direction, restricted by a rotation threshold. The alignment with the neighbors is done by steering the direction towards the \textit{average heading} of neighboring birds. To do this we implemented the angular average with the formula \\ $\bar \alpha = atan2 \left( \frac{1}{n} \sum_{j=1}^n sin(\alpha_j),\frac{1}{n}\sum_{j=1}^n cos(\alpha_j) \right)$ with $n$ the neighboring birds and $\alpha_j$ the neighboring bird's heading. We limit the alignment turn with a threshold parameter. Finally for the cohesion rule, we turn the direction of the bird towards the center of mass of local flock-mates.

To move a bird forward in the advance function call, we calculate the $x$ and $y$ increments by decomposing the heading vector into its components. At the transition, the new position is informed to the coupled model by means of an upward causation message, which will in turn trigger the $\delta_G$ function that computes the nearest neighbors for each bird. 
 
The value coupling function serves the Birds with information regarding the closest bird's distance and heading, the flock-mates center of mass and average direction, and the overall number of flocks in the system. The later is used in two extensions of the Boids model discussed later in this section. This information is used in the aforementioned $\dint$ closing the circle for the closed-loop dynamic.

Extending this model provides us with the possibility to experiment with different closed-loop feedback dynamics. Two different extensions have been developed using emergent macro-level information to change micro-level (bird) behavior.

\textit{Fearful agents avoid gangs} (or FA) presents a modification in the
internal transition of the bird's models. When the number of flocks at  the
system surpasses a defined threshold, the agents change their behavior.
Instead of cohering and aligning with the neighboring birds, they switch to 
an anti-cohesion behavior. This behavior changes headings to point towards the
opposite direction of the closest neighbors' center of mass. This will apply while the number of flocks remains above the threshold. The number of flocks is calculated in the $\delta_G$ and informed in the $v_{down}$ function.

\textit{Brave agents join gangs and develop laziness} (or BA), like FA, 
takes into account the total number of clusters of the system. Yet, instead
of doing anti-cohesion, it will make the birds drive their headings towards the center of mass.
We call this super-cohesion. This process will happen for a fixed amount of iterations and will occur for a
limited number of times during the simulation. The length of each period will
decrease for each iteration the bird enters this state and it will be enabled a limited amount of times.

\subsubsection{Simulation Experiments} ~ \\

The experiments detailed in this section were configured with the following parameters: the size of the grid is 70 x 70 units, the flock size is 200 birds, the radius of
visibility for each bird is 5 units and the minimum allowed distance between birds is 0.5 units. The position evolves over time following the updated direction with a constant velocity equal to 1 distance unit per time unit. The total simulation time was set to 250 time units.

\begin{wrapfigure}[19]{r}{0.5\textwidth} 
\begin{center}
    \includegraphics[width=0.48\textwidth]{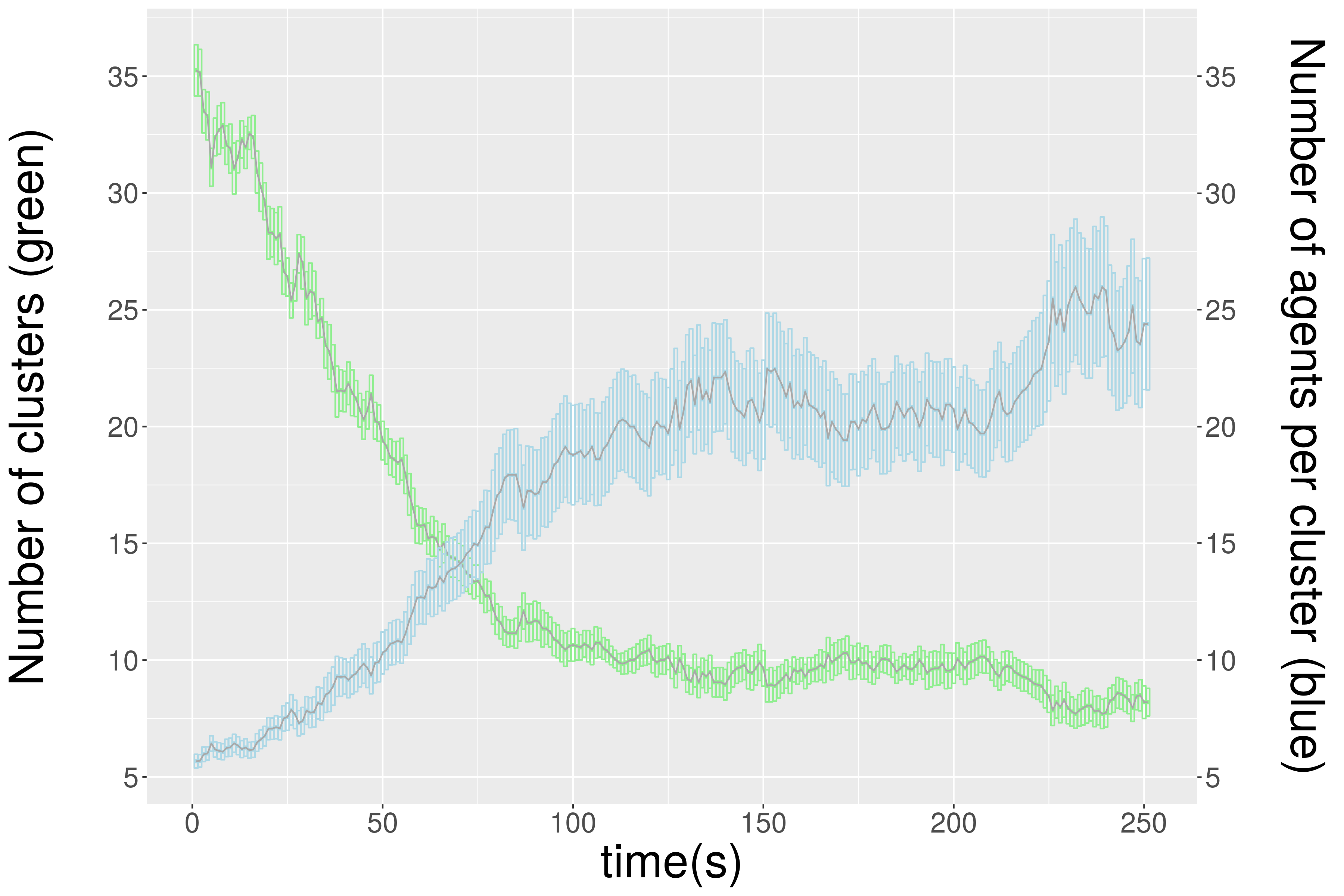}
    \caption{Boids vanilla model. X-axis shows time, the green curve shows the evolution of mean and standard deviation for the amount of clusters. In blue, the evolution of the mean and standard deviation of the number of agents per cluster.}
    \label{fig:boids_vainilla}
\end{center}
\end{wrapfigure}

\begin{figure}[!bp]
  \begin{subfigure}[b]{0.5\textwidth}
    \begin{center}
      \includegraphics[width=\textwidth]{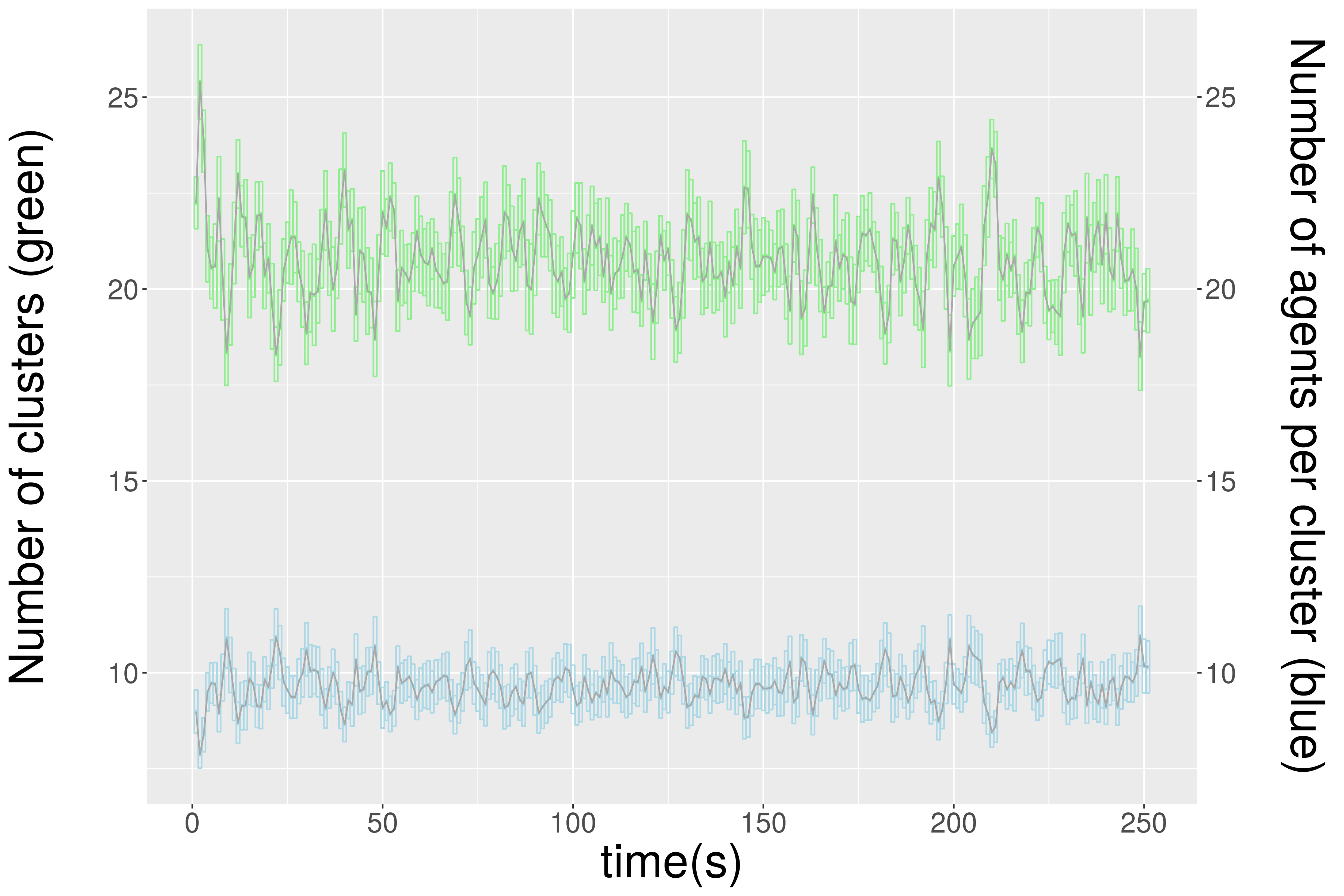}
      \caption{Boids FA model. }
      \label{fig:boids_faav} 
    \end{center}
  \end{subfigure}
~
  \begin{subfigure}[b]{0.5\textwidth}
  \begin{center}
    \includegraphics[width=\textwidth]{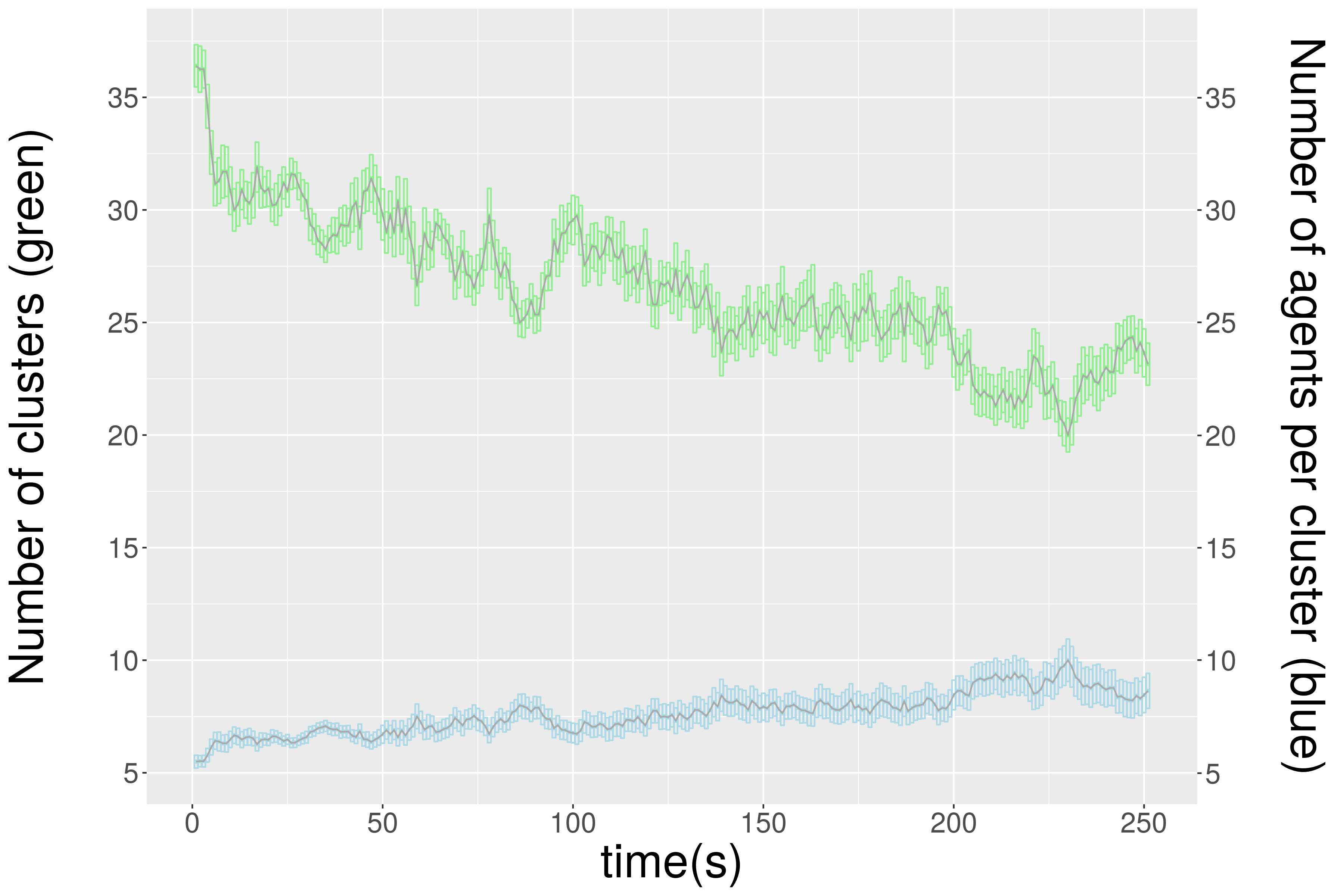} 
    \caption{Boids BA model.}
    \label{fig:boids_baaj}
  \end{center}
  \end{subfigure} 
  \caption{Boids models clusters evolution. X-axis shows time, the green curve shows the evolution of mean and standard deviation for the amount of clusters. In blue, the evolution of the mean and standard deviation of the number of agents per cluster.}
\end{figure}

In the following plots we show the average and standard deviation of the amount of clusters and the number of agents per cluster. These experiments shows 20 simulation runs aggregates.
The \autoref{fig:boids_vainilla} shows the original Boids model simulation, \autoref{fig:boids_faav} shows the modified Fearful Agents model simulation. In \autoref{fig:boids_baaj} we see the Brave Agents model simulation.

\begin{figure*}[!tp]
    \centering
    \includegraphics[width=0.85\textwidth]{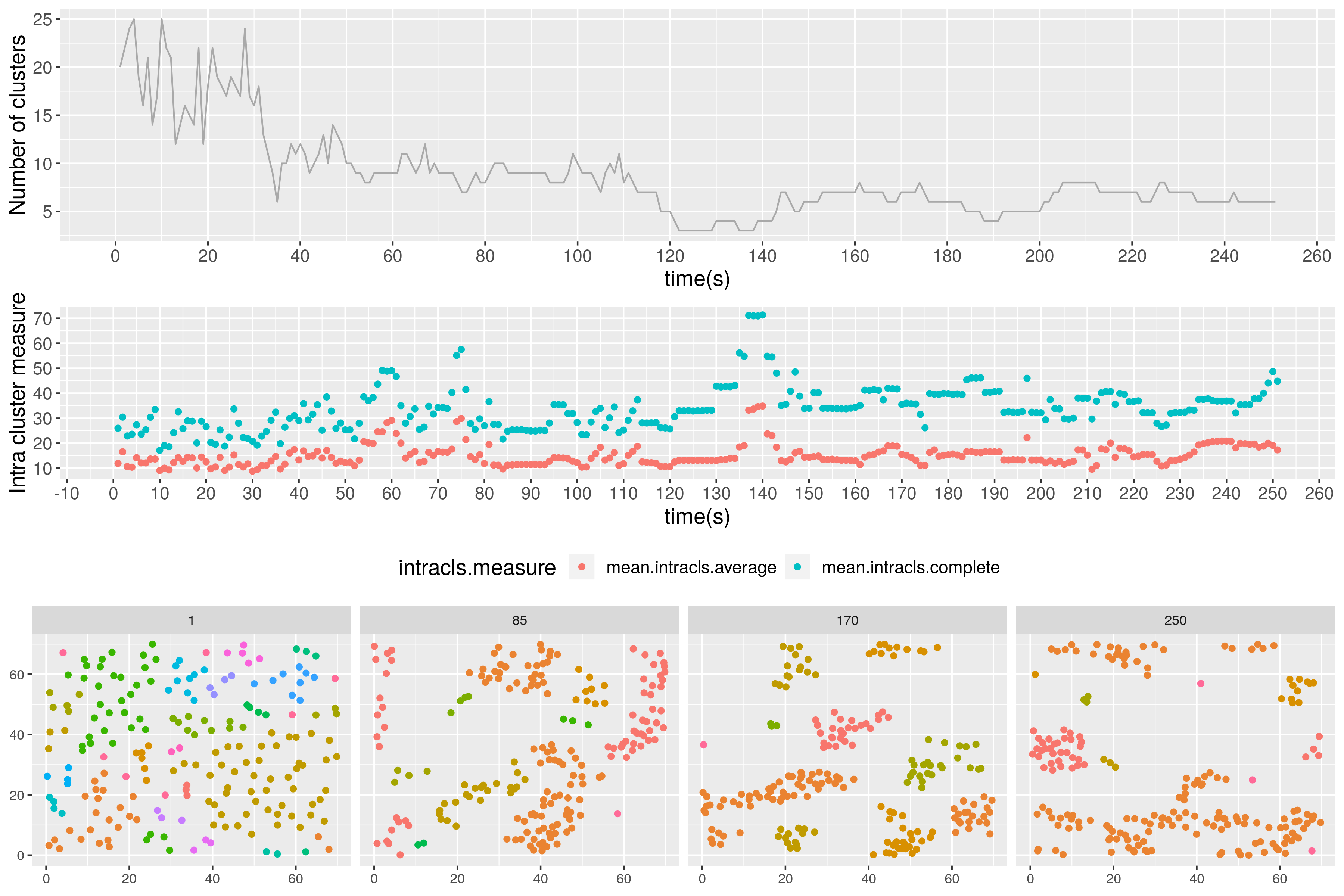}
    \caption{Boids Vanilla model. 
    Top sub-figure shows the evolution of the amount of clusters. Middle sub-figure shows the intra-cluster distance. Bottom sub-figure shows the positions of the birds in the 1st simulation second, 85th second, 170th second,  and 250th second (the final state of the simulation). Bird's color represent the flock membership, same color same flock. }
    \label{fig:boids_vanilla_single}
\end{figure*}

\begin{figure*}[bp]
    \centering
    \includegraphics[width=0.85\textwidth]{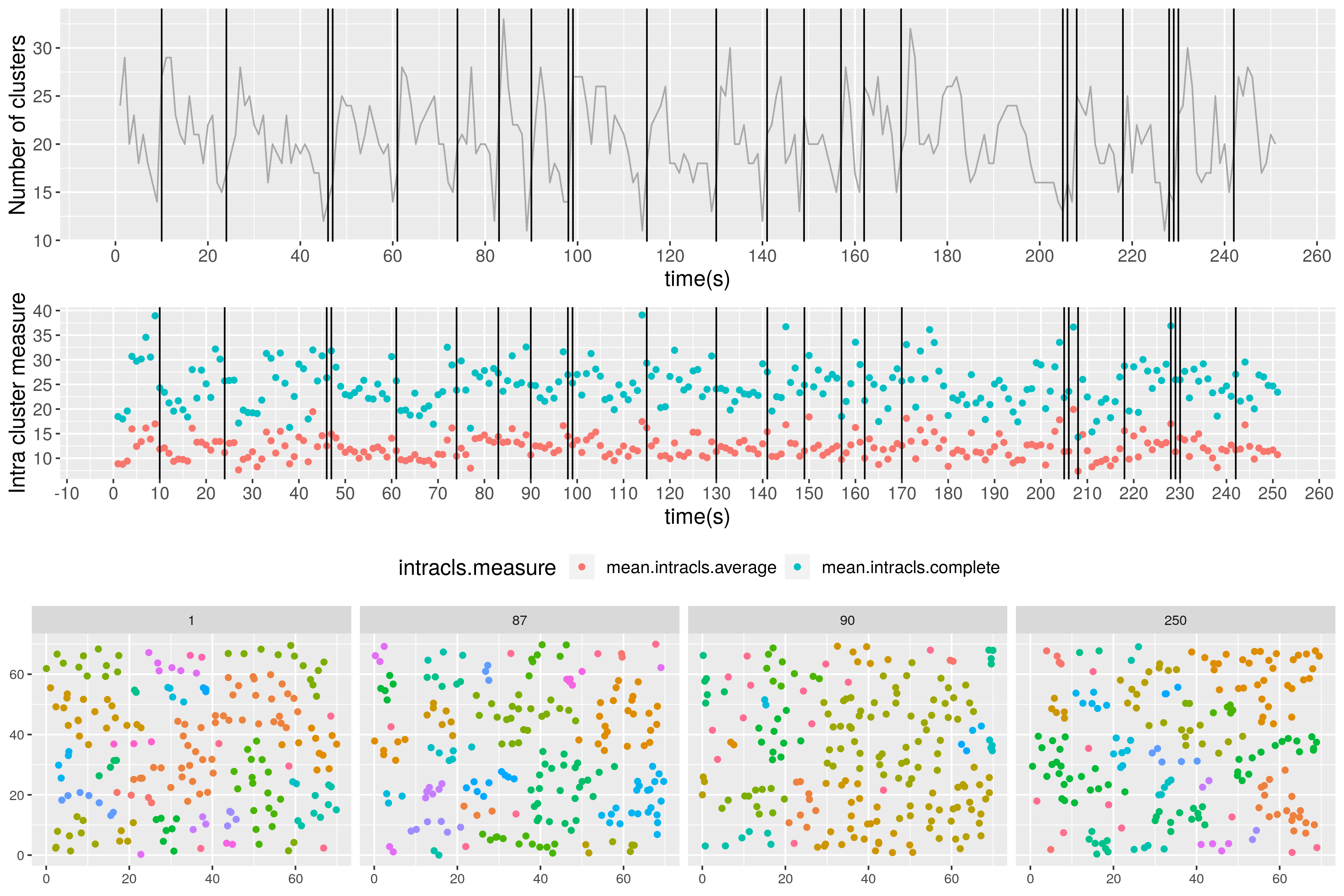}
    \caption{Boids FA model. Top sub-figure shows the evolution of the amount of clusters, vertical black lines depict the moments where the anti-cohesion event was detected. Middle sub-figure shows the intra-cluster distance. Bottom sub-figure shows the positions of the birds in the 1st simulation second, 87th second  (before anti-cohesion event), 90th second (after anti-cohesion event), and in the 250th second the final state of the simulation.}
    \label{fig:boids_fa_single}
\end{figure*}

\begin{figure*}[!ht]
    \centering
    \includegraphics[width=0.85\textwidth]{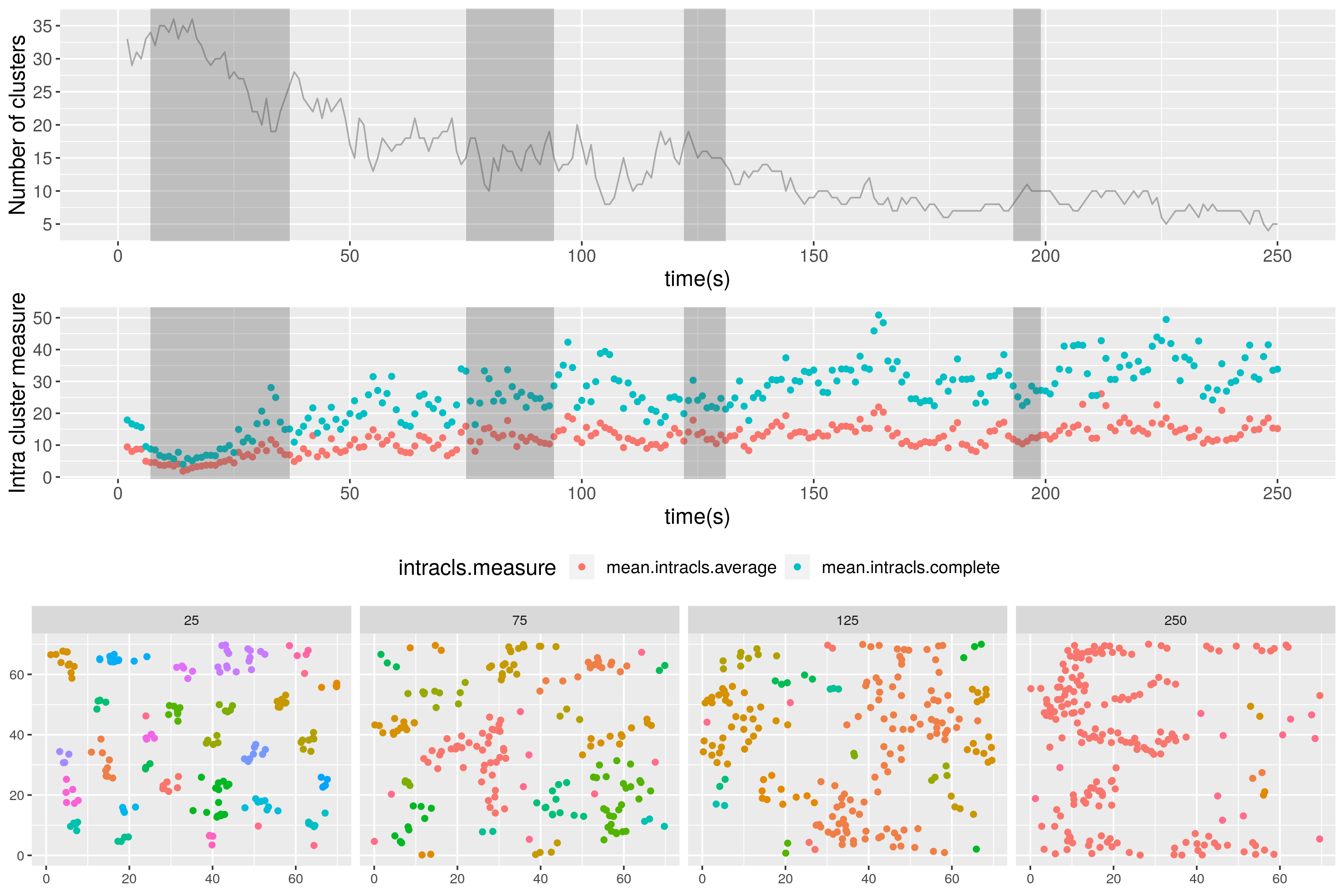}
    \caption{Boids BA model. Top sub-figure shows the evolution of the amount of clusters, gray boxes depict the super-cohesion event. Middle sub-figure shows the intra-cluster distance. Bottom sub-figure shows the positions of the birds in the 
     1st simulation second, 25th second (first super-cohesion event with minimum intra-cluster distance), 75th second (second super-cohesion event), 125th second  (final super-cohesion event), and 250th second (the final state of the simulation).}
    \label{fig:boids_ba_single}
\end{figure*}

The effects of the macro events affecting the micro level agents are better appreciated in single-run experiments. We show in the figures \autoref{fig:boids_vanilla_single} (Vanilla model), \autoref{fig:boids_fa_single} (FA model), and \autoref{fig:boids_ba_single} (BA model), the evolution of the amount of clusters (top), the intra-cluster complete and average distance (middle), and the agents evolution on the grid (bottom) for each model.


\subsubsection{Discussion}~\\

In this section we presented and extended the classic Reynolds' Boids model to demonstrate the use of closed-loop dynamics. We provided two extensions implementing different scenarios of micro-macro interactions.

The EB-DEVS' Boids implementation, shows how the formalism assists the modeling phase by providing tools that simplifies the coding of the Boids rules into the model. Using dynamic structures like the Radius Neighbors Regressor, allowed us the identification of emergent structures in real time. The bird to bird interactions developed complex behaviors that resulted in higher level structures if observed from the environment point of view. These cluster-like objects influence the birds behavior showing group cohesiveness.

Let us first consider the Vanilla Boids model implementation. We were able to observe that for every experiment run, the number of flocks (or clusters) decreases while growing in size. The generation of large flocks, and the convergence of the birds flocks into well organized structures is observed in every run.  In  general, this behavior is followed by an increase in the size of the flocks (\autoref{fig:boids_vainilla}). Moreover,  when we start considering the flock as a higher level structure, we can observe properties of its own. The intra-cluster diameter is one of them and it can be used to analyze flock cohesiveness. During the single run we were able to analyze this property related to the number of  system level clusters. Furthermore, the number of clusters drop while the complete and average intra-cluster grows. These measurements help to understand the flock's cohesive behavior (\autoref{fig:boids_vanilla_single}).

In the case of the FA model, we observed how the anti-cohesion rule resets the flock structure. The spontaneous disorder generated by this rule, is enabled by a system level property. This information triggers birds'  anti-cohesion behavior. In this scenario, the model presents no adaptation in the birds' behavior. In opposition to the vanilla model, it is noteworthy to show how convergence is never reached (\autoref{fig:boids_faav}). Furthermore, the number of clusters and their average sizes spikes shortly after the anti-cohesion rule. Additionally, the analysis of intra-cluster distance shows how cohesion is far from being present  (\autoref{fig:boids_fa_single}).

Finally, in the BA we represent  adaptation and self-organization thanks to the super-cohesion rules. During the execution of the model we can see an increase in the number of flocks during the first moments of the experiment's execution. This is explained by the initial super-cohesion behavior. Uniformly distributed Birds tend to gather up in very cohesive flocks (\autoref{fig:boids_baaj} and \autoref{fig:boids_ba_single}).  After the initial phase, the birds  form new flocks with increasing inter cluster distance. Furthermore, this behavior is like the one observed in the Vanilla model. While the adaptation of the super-cohesive events decreases, the behavior tends to reproduce the Vanilla model.

In summary, the Boids model allows us to put in place different micro-macro level interaction patterns. With different behavioral rule-sets, we can implement a diverse set of distributed behavior algorithms. Furthermore, we can design and experiment with complex system's adaptive, self organizing behaviors.

\subsection{Agent-Based Modeling of Mitochondria Sub-Cellular Dynamics} 

Mitochondria are highly dynamic organelles placed within eukaryotic cells. They are constantly undergoing fusion and fission events to adapt to stress conditions and  to the cell's energy demand. The dynamics and homeostasis of this sub-population was studied by Dalmasso et al.\ \cite{Dalmasso2017Agent-basedHeterogeneity} by means of a simulation model, where each mitochondrion moves in a 2D plane limited by an outer circle (the cell membrane) and an inner circle (the cell nucleus). In the space between these two borders each mitochondrion moves with a random velocity vector at discrete time steps $\Delta$t = 1s. After 300 seconds the mitochondrion can undergo a fission with a predefined probability, or a fusion if a suitable mitochondrion is nearby. We slightly modified the model proposed in \cite{Dalmasso2017Agent-basedHeterogeneity} by restricting fusion and fission cycles to be mutually exclusive, based on the work of Twig et al. \cite{Twig2008FissionAutophagy}.

\subsubsection{Mapping to EB-DEVS}~\\

The Mitochondria model is a discrete time model that uses indirect communication between mitochondria across the cell. The mitochondrion is represented as atomic models.
The environment is a coupled model named Cell that defines three regions:
nuclear, perinuclear and cytosolic areas.
Atomic models track the movement of each mitochondrion  through perinuclear and cytosolic regions, considering that they cannot enter the nucleus or escape the cell.
The mitochondrial speed is determined by a uniform random value. In the perinuclear area, mitochondria move on average slower than in the cytosolic area.
The Cell presents certain mass restrictions, such as the smallest and largest possible mass of a mitochondrion.

We now show the $\dint$ definition in \autoref{lst:intmito} describing how the model is implemented.

First, the $\dint$ transition function changes the $(x, y)$ coordinates by moving the mitochondrion in the new direction by projecting the heading vector into $(x,y)$  coordinates.
If the cellular boundary is reached, the direction is set to  the opposite direction.

The fusion and fission cycles take effect every 300 seconds, thus the $\dint $ transition checks if it is time for a fusion/fission cycle to take place (\autoref{lst:intmito} lines 6-7).

For a mitochondrion to fuse with another, several conditions need to be met (\autoref{lst:intmito} lines 13-15). A uniformly distributed number is generated and compared with the predefined fission probability to determine if it will fission.
If this condition is met, the mitochondrion must have enough mass to split into two mitochondria, each one with at least the minimum size (in this model 0.5 $\mu m^2$).
As the fission event proceeds, an inactive mitochondria becomes active taking its remaining mass and positioning at the same coordinates as the original (\autoref{lst:intmito} lines 29-32). Hence, the system preserves mass during fusion and fission cycles.

The fusion of two mitochondria also takes place every 300 seconds in a similar fashion.
First, a uniformly distributed number is generated and compared with the predefined fusion probability.
If the condition is met, there needs to be a neighboring mitochondrion such that the sum of both sizes is less than the defined maximum size (in this model 3 $\mu m^2$) (\autoref{lst:intmito} line 15). In the meantime, the fused mitochondrion gets the neighboring mass and the neighboring mitochondrion is set to \textit{'inactive'} (\autoref{lst:intmito} lines 33-35).

Fusion and fission events' probabilities are disjoint, in opposition to Dalmasso's model. Yet, the case where a mitochondrion remains unchanged during the fusion-fission cycle still exists. Furthermore, we modified Dalmasso's model to behave in this manner for comparison purposes. In summary, we used a modified version of the NetLogo model with disjoint fusion and fission events for comparison purposes.

As we defined EB-DEVS with static couplings in mind, agents cannot be created or removed from the coupled model's domain in runtime. Thus, we implemented a mechanism to activate/deactivate (instead of add/remove) atomic models from the cellular dynamics at runtime. Our model features  \textit{'inactive'} and \textit{'active'} atomic models. An atomic \textit{inactive} model becomes \textit{'active'} if a new mitochondrion is created by a fission event.
\textit{Active} agents go through motion and fusion-fission cycles, \textit{inactive} agents wait for activation signals.

After each internal transition, the new model states are communicated to the Cell via the $y_{up}$ function. This communication allows the coupled model to synchronize the list of active and inactive models.
The coupled model informs the atomic models with restrictions based on the area of the cell it is located at, such as proximity with other mitochondria. It also enables for structural changes by keeping an updated list of active and inactive models.

\begin{lstlisting}[float,style=base,
linewidth=\columnwidth,breaklines=true,
    basicstyle=\ttfamily\footnotesize,
    xleftmargin=2em,
    numbers=left,
    multicols=2,
    stepnumber=1,
    tabsize=2,
    firstnumber=1,
    numberstyle=\tiny,
    numberfirstline=false,
    label={lst:intmito},
    caption={Pseudocode of the Mitochondrion state internal transition}]
# $F_p$ = Fusion probability
# $f_p$ = Fission probability
# $M_{MAX}$ = maximum mass allowed
# $M_{MIN}$ = minimum  mass allowed
function intTransition:
  fusion_fision_condition = current_time mod 300 == 0
  if not fusion_fision_condition:
    if (model is Active):
      calculate_area_from_position()
      calculate_velocity_form_position()
      advance()
  else:
    if (mode is Active and
      unif(0, 1) <= $f_p$ and
      mass $\geq 2* M_{MIN}$):
        x_f = unif(0, 1)
        m_1 = $(x_f * (0.5 - \frac{{M_{MIN}}}{mass})$ + $\frac{{M_{MIN}}}{mass}) * mass$
        old_mass = mass
        mass = m_1
    if (mode is Active and
      unif(0, 1) < $F_p$ and
      mass $\leq 2* M_{MIN}$):
        closest_mito = get parent closest active
                        neighbor information
        total_mass = mass + closest_mito.mass
        if total_mass $\leq M_{MAX}$:
          mass = total_mass
          inform parent to inactive closest_mito
    if (state is Inactive):
      fission_model = get parent fissionated model information
      if fission_model:
        mass = fission_model.old_mass - fission_model.mass
        state = Active
      should_fusion = get parent fusion information
      if(should_fusion):
        state = Inactive
  inform to parent new state
  return state
\end{lstlisting}

\subsubsection{Simulation experiments}~\\

\begin{figure*}[!tbp]
    \centering
    \includegraphics[width=\textwidth]{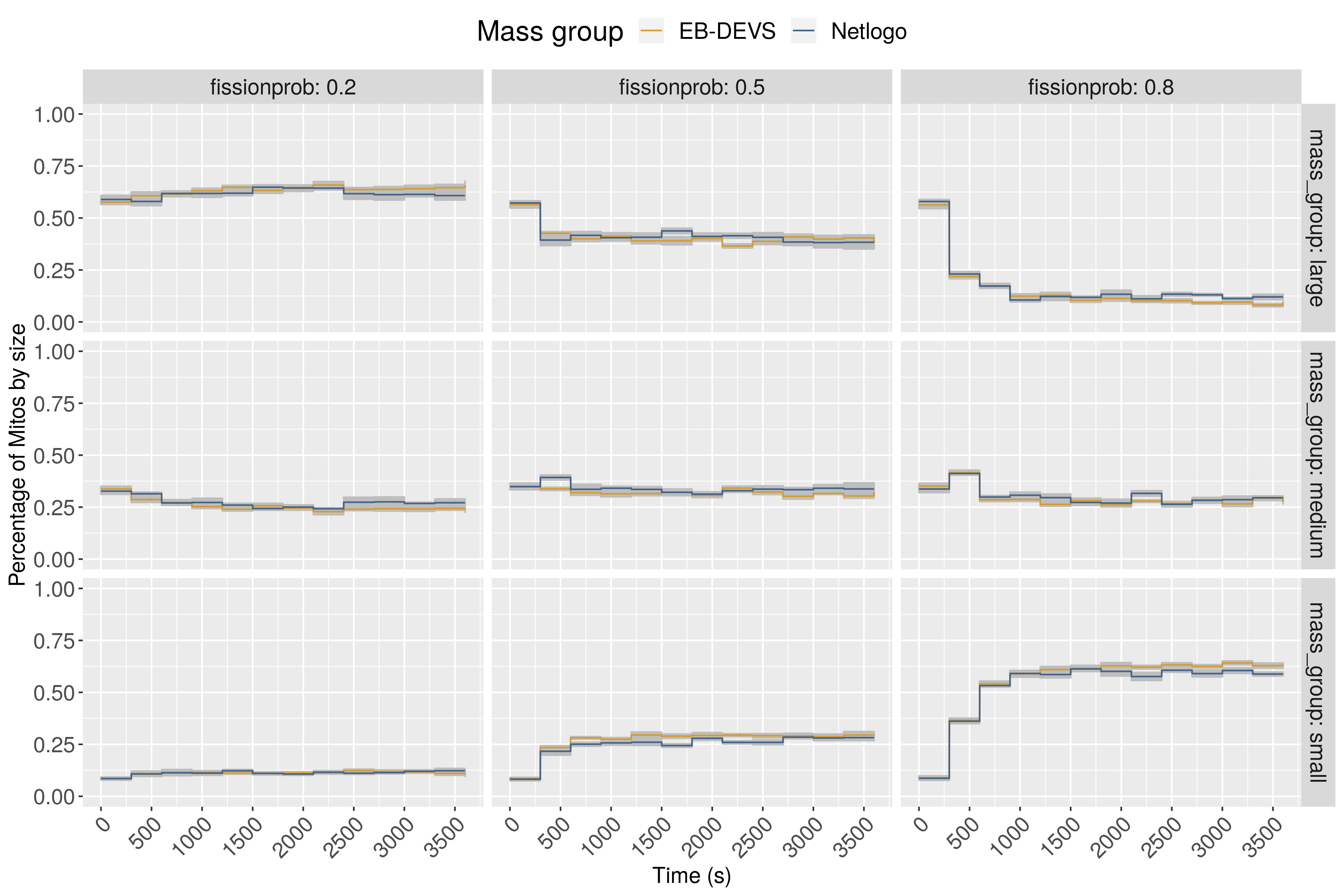}
    \caption{A comparative between Dalmasso's (modified) mitochondria model and the EB-DEVS implementation. The x-axis presents the evolution in time. y-axis shows the percentage of mitochondria for each size, small medium and large. Mitochondrial mass groups:  $0.5\mu m^2 \leq small \leq 1\mu m^2  < medium \leq 2\mu m^2  < large  \leq 3\mu m^2$. Three fission probabilities were tested according to Dalmasso's results.}
    \label{fig:mito}
\end{figure*}

We validated the implementation of the model against a  modified Dalmasso's model implementation.
Both models were initialized with uniform distributed mitochondrial sizes and a total size of 300 $\mu m^2$.
We run 3 different combinations of fission/fusion probabilities (20\%/80\%, 50\%/50\% and 80\%/20\%). 
Since both approaches are stochastic, we ran each model 20 times for 1 hour of virtual time, which corresponds to 12 fusion/fission events.
The results in \autoref{fig:mito} show that the system is able to reach the state of homeostasis. 
In blue we see the evolution of the NetLogo version, and in orange the EB-DEVS implementation.
The mitochondrial sizes are grouped as in the Dalmasso's model:\\
\begin{align}
  0.5\mu m^2 &\leq small   \leq 1\mu m^2 \\
  1\mu m^2 &< medium \leq 2\mu m^2\\
  2\mu m^2 &< large  \leq 3\mu m^2
\end{align}
Each mitochondrion falls in one of these size-based groups.

\subsubsection{Discussion}~\\

In this section we provided a model that captures mitochondrial dynamics. 
The implementation of Dalmasso's mitochondrial model using EB-DEVS allowed us to create closed-loop dynamics with multiple types of event cycles. Movement, fusion and fission cycles alter the system's structure and mitochondria's behavior. Such behavior is responsible for the consequent system stabilization; after three to four fusion fission cycles the mitochondria's group distribution converges towards a stable value.
The authors of the original model observed this phenomenon in the cellular sub-populations of different masses, calling it \textit{sub-population homeostasis}. This property emerges from the lower level models towards the higher levels of observation of the system. This can be considered an emergent property unforeseen from the agents individual behavior.

The validation of the model was done against the modified Dalmasso's NetLogo model to restrict for mutually exclusive fusion and fission cycles. The results of the executions show good 
correspondence between models, either while comparing means, standard deviations, and overall behavior.

Considering the level of complexity of this model, we found that it tests the limits of the EB-DEVS approach. In the first place, fusion and fission cycles use the environment while sharing mass in the fusion and fission cycles. We used indirect communication for the implementation of this model's feature. To our knowledge, without indirect communication, this feature would require either dynamic input-output ports connections or clique connectivity between atomic models. Using dynamic input-output ports would also be required to represent complex behavior as the links should be updated based on mitochondria's proximity. While DEVS extensions like ML-DEVS ~\cite{Steiniger2016} or Dyn-DEVS \cite{Uhrmacher2001} include dynamic links as part of the formalism, they fall far away from the EB-DEVS approach in several other aspects. 
We shall tackle dynamic structure in a follow up paper extending EB-DEVS. Additionally, another benefit from using indirect communication lies upon the model's declarability and performance. Let us consider the case of using a model with \textit{n} agents. In this case, 
the number of communication links (connections between input and output ports in the model) would be $n(n-1)$, most of them remaining actually unused.
Modeling a full mesh of mostly unused links adds unnecessary clutter that can hinder both performance (e.g. memory footprint) and model readability. This is avoided by the indirect communication that takes place naturally in EB-DEVS via micro-macro feedback loops.

Another challenging modeling problem can be seen in the dynamic nature of mitochondria. The model of mitochondria requires that mitochondria are created and destroyed in runtime. We were able to model this using active and inactive models. 
At the moment a fission event occurs, an inactive agent turns into an active one. 
While this strategy works, it penalizes the simulation performance. 

By using these inactive models, we doubled the required models for the system to work.

In summary, our model implementation reproduces Dalmasso's emergent behavior of sub-population homeostasis while preserving total cellular levels of mitochondria. This behavior emerges out of the interaction between macro and micro-level. Here, the environment plays a leading role in allowing indirect communication between mitochondria. The environment assists fusion and fission cycles across three types of mitochondria populations, cycles responsible for size changes while traversing the cellular space.

\section{Discussion and related work} 
\label{sec:discussion}

In this work we have discussed the new EB-DEVS formalism that focuses on dealing formally with micro-macro closed-loop dynamics.

Emergence is a niche research area that has many interesting applications. There is a growing need to develop techniques for the identification, validation and modeling of emergence. 

Our intention is to highlight the importance of having a formalism that helps the modeler with expressing emergence in complex systems. 

\subsection{EB-DEVS compared to other approaches}
\label{sec:disc_comp}

We can find in the literature several efforts to tackle the identification and validation of emergent properties. A review of the state of the art can be found in Szabo's and Birdsey's work \cite{Szabo2017}. The authors describe  critical challenges in the identification of variables capturing micro and the macro levels, and the relationships and dependencies between them. Correspondingly, the modeling and simulation discipline's needs to address this with tools for the validation and identification of emergence in simulated models.
Regarding emergent properties identification, the authors classify the efforts in postmortem vs live, depending on whether it is done after the system's execution or while it is running respectively.


Regarding DEVS and emergence, Bernard Zeigler studied DEVS's ability to model and predict \textit{emergence} \cite{Zeigler2016a}. Zeigler's work shows how different formalisms based on DEVS allows the analysis and prediction of \textit{emergence}. He identifies three key aspects for the modeling of emergence. In the first place, the importance of how dynamic coupling mechanisms enables to model complex systems  and how the appearance of new technology facilitates the modeling of such evolving systems. Also, Zeigler argues in favor of Markov chain DEVS \cite{seo2018devs} for their explanatory and predictive power. Finally, Emergence Behavior Observers~\cite{mittal2015harnessing} (EBO) can be useful as a strategy to analyze system state transitions with the purpose of recording model's states snapshots.

Recently, Cellular Automata (CA) has been considered as a paradigm of emergence in the complex systems field \cite{Chalmers2011}. We can see an example of this in the emergent patterns of Conway's game of life. In this sense, we can mention Gabriel Wainer's CellDEVS \cite{Wainer2002}, a DEVS Extension that integrates Cellular Automata with DEVS simulation models. The idea is to bring the benefits of DEVS formal simulation models to the CA world. CellDEVS shows its applicability in diverse scenarios  \cite{Wainer2001,Ameghino2001,Wainer2010a,Muzy2002}. It allows for indirect communication between neighboring cells and it results in a sound tool to model spatially explicit complex dynamics. Yet, it depends on a predefined fixed neighboring region for each cell, limiting the capability of producing  emergent behavior in more generalized situations. 
Also, CellDEVS does not provide a means to discover and handle explicitly system-level emergent properties.


Regarding dynamic structures in DEVS, Barros \cite{Barros1997} defines DSDEVS a formalism where its network structure can be modified in runtime. Barros extends DEVS with the use of a particular type of model, the \textit{network executive} (NE), a hybrid  atomic/coupled model responsible for network changes. The NE's state changes via transitions and presents a \textit{structure function} that changes the network structure according to its own state. It can be argued that the network structure emerges from the dependant's states and that this network change influences back the agents. Nevertheless, the lack of a closed-loop feedback mechanism restricts the possibility of modeling emergence, while the required hybrid atomic/coupled model can be seen as an entity too artificial when comparing the model against the real system.

       
Steiniger and Uhrmacher introduced ML-DEVS, a general purpose DEVS extension that combines ``\textit{a modular, hierarchical modeling with variable structures, dynamic interfaces, and explicit means for describing up and downward causation between different levels of the compositional hierarchy}' \cite{Steiniger2012}. ML-DEVS presents several differences with Classic DEVS\@. First, the formalism presents a different model's taxonomy. Coupled models have their own autonomous behavior, like classic Atomic DEVS models. The top-down $\lambda_{down}$ function, activates bottom models in a downward causation fashion, the upward information and causation are modeled using variable ports. Changing the port's distribution generates changes that enable structural changes and upward causation. Even though ML-DEVS allows for closed-loop dynamics via downward and upward causation/information, it requires definitions of atomic and coupled models that are considerably different from Classic DEVS (e.g. atomic models have a single transition function, the coupled model has its own time advance function, to name a few).

This hinders a smooth integrability between Classic DEVS and ML-DEVS within a same simulation model, which is one of the main goals pursued by EB-DEVS (for a full implementation of ML-DEVS we refer the reader to the James II implementation and corresponding publications \cite{Himmelspach2007PlugnSimulate}).

All previously described DEVS extensions provide tools that target indirect agents' communication, multi-level feedback loops, dynamic structures, and emergence prediction and modeling. Nevertheless, our contribution provides a general modeling framework for the live-identification and validation of emergent properties compatible with existing DEVS simulators.

There exist other formalisms beyond the DEVS ecosystem that support multi-level modeling, and therefore could be valid alternatives to deal with live-identification and modeling of emergent properties. Examples are ML-RULES\cite{helms2017mlrules} (
a modeling language for the modeling of biological cell systems, centered in the description of rules supporting hierarchical dynamics, assignment of attributes at each level and flexible definition of reaction kinetics),
Coloured Petri Nets\cite{liu2019coloured} (a Petri Net's extension specific for the modeling of complex systems enabling multi-level, multi-scale and multi-dimensional models),
and Attributed $\pi$-calculus with Priorities\cite{uhrmacherPiCalculos} (an extension of the $\pi$-calculus formalism for the modeling of concurrent systems with process attributes, enabling interaction constraints and expression of multi-level dynamics),
just to name a few.

There are two other contributions to the live identification of emergent properties that we consider worth mentioning. Szabo and Teo have developed an approach for the semantic validation of emergence~\cite{szabo2012objective}. This approach is used either for a-priori or postmortem identification and validation of emergent properties. Identification is done by computing the semantic distance between state variables and the model’s composed-state. Using their modification of reconstructability analysis it is possible to calculate such composed-state and, if the semantic distance is significant, to save this state as an emergent state. In the context of EB-DEVS this technique can be implemented at the new global state transition function of the parent (coupled) level.

Chan's work \cite{Chan2011a} analyses interaction metrics between agents. He gives arguments in favor of this approach for the detection of emergent behavior.

Regarding taxonomy, different authors have categorized emergence identification in three classes, grammar-based, event-based, and variable-based.

In event-based emergence modeling (see Chen's work \cite{Chen2009}) emergence is defined as complex events that can be reduced to a sequence of simple events. Simple events are defined as state transitions and complex events are a composition of complex or simple events. Chen used event-based modeling for postmortem analysis.

Kubik \cite{Kubik2003} proposed a grammar-based approach towards the identification of emergence. It relies on the idea that ``\textit{the whole is more than the sum of its parts}'' and specifies two grammars, $L_{WHOLE}$ and $L_{PARTS}$. The former grammar represents the system properties, while the latter represents the properties resulting in the sum of the parts. The difference between $L_{WHOLE}$ and $L_{PARTS}$ results in the properties that are found in the system and not in $L_{PARTS}$, these are emergent properties. This approach is used mostly for the postmortem identification of emergent properties. Szabo and Teo have extended these ideas \cite{Szabo2015} to support multiple agent types, mobility and variations on the number of agents over time. 
        
Several efforts in emergence identification can be found in the variable-based category. The use of a variable or metric for identification of changes in the `expected' behavior of the system. Some authors like Mnif \cite{Mnif2006}, use entropy to measure emergence. As entropy decreases in the selected variable, the signs of order appear showing evidence of `self-organization'. It would be interesting to explore joint-entropy using multiple variables in this analysis. In this line of research, Seth's work \cite{seth2008measuring} uses Granger's causality to establish relationships between the model's levels, this approach requires stationarity for correlated variables. This property is hard to determine without the complete observation of the variables interactions, hence the methodology is best suited for postmortem analysis. Furthermore, using multiple variables for Granger's causality presents its own problems like multicolinearity.

In addition, some authors used Machine Learning for the identification of emergent properties. Brown and Goodman \cite{Brown2014} used naive Bayes to classify the collective behavior of a flock model depending on the spatial structure and neighboring interactions between birds. Via monitoring the trajectories of stochastic simulations, statistical model checking ~\cite{agha2018} allows to test whether (possibly emergent) properties specified in a temporal logic occur with a specific probability or to determine the probability with which these properties occur.

We claim that our EB-DEVS approach is compatible with all the above listed ongoing research efforts, towards the live identification of emergent properties that rely on metrics based on interactions, behavior or properties. We argue that EB-DEVS can be seen as a \textit{container} where such strategies can be implemented and tested, and that our main contribution relies on a generic framework towards live identification of emergence based on DEVS.


\subsection{Emergence and complex systems' modeling}
Emergence is a broad topic that has been discussed in the literature for over a century. Philosophy and science have treated this subject extensively, as it is intertwined with the interpretation, understanding and analysis of complex systems. Either if we study life, consciousness, or engineering systems, emergence is a key element for their understanding. Emergence is relevant while analyzing a system's properties, and the underlying patterns and first principles that drive a systems' behavior. 

As with many complex terms, there are many definitions for emergence.  We would like to give here some useful definitions while working with complex systems and emergent behavior. Let us start by defining emergence according to three modern authors. 

John H. Holland \cite{Holland2006} defines emergence in a system \textit{when} the sum of its parts is bigger than the whole, giving place to non-linear behaviors. Baas \cite{Baas2018} defines a property as emergent, if it is discontinuous from the properties of the components at the lower levels in the hierarchy.
While Fromm \cite{Fromm2006}  defines it as the formation of order from disorder that is based on self-organization, or as the materialization of higher-scale properties unexpected while observing lower-scale behaviors.

From these definitions we can identify some common characteristics. The notion of a hierarchy of components that connects the system's at different scales is present, the components show properties that in a higher level of the hierarchy present different characteristics than in isolation. Finally, the formation of higher level properties that shows levels of organization in the lower level components.
        
But how do these different levels connect, what drives self-organization, and what are their relationships with simulation? Several authors have discussed these issues in the past. There is broad agreement about at least two types of emergence: \textit{Weak} and \textit{Strong}.

Mark Bedau highlights the importance of weak emergence over strong emergence \cite{bedau1997weak}, stating that ``at best [strong emergence] play[s] a primitive role in science. Strong emergence starts where scientific explanation ends'' (Bedau, 1997, p.7). While deciding which type of emergence is important for science, and in particular for the M\&S discipline, we should favor weak emergence as it allows the study of the micro-level mechanisms that may be causing the emergent phenomenon.
    
A definition for weak emergence that relates to multi-level dynamics is described by Szabo and Birdsey, which we choose to adopt for our work:  ``\textit{weak emergence as being the macro-level behavior that  is a result of micro-level component interactions, and strong emergence as the macro-level feedback or causation on the micro-level}'' (Szabo, 2017, p.4)~\cite{Szabo2017}.

As for multi-level system's interactions Campbell\cite{campbell1974downward} and Emmeche \cite{EMMECHE1997} among others have discussed how higher levels affect or restrict the course of action of bottom level processes. Downward causation, a term coined by Campbell, reflects how the system at the top-levels affects the bottom levels. Upward causation can be seen as a reciprocal relationship where bottom-level entities affect the top-level. Apparently, these two notions cannot be separated. Mario Bunge's treat on causality \cite{bunge2017causality} (p.362) exposes both relationships and criticizes the use of the term `causation' in this context, as he states that ``\textit{what we do have here is not causal relations but functional, relations among properties and laws at different levels}''.

Evidently, the causation or functional relations between levels are intertwined with weak and strong emergence.

In addition to these concepts that relate the micro with the macro, it is important to discuss how system components at the same level relate and communicate with each other. DEVS adopted a direct communication strategy. For each pair of models we can define an input output port's bond that allows for direct message passing. Analogously, indirect communication allows for the interaction between models without the need of an explicit communication channel.

The role of communication in multi-agents system's simulation, either direct or indirect, and its evolution, has been well documented. For instance, in \textit{Multi-Agent System Simulations} Tummolini et al. \cite{Tummolini2009StigmergicApproach}, discusses the role of communication and its evolution in the discipline. In particular, the role of stigmergy in simulation models.

Stigmergy is a form of indirect communication through the environment where the actors or components use information put at disposal by an upper-level layer in the system. The concept was proposed by Grassé \cite{Grasse1959LaConstructeurs} while observing the organization of termite populations. Stigmergy was further  conceptualized by Mittal \cite{Mittal2013} in the context of the DEVS formalism.

In the case of EB-DEVS, we introduced changes in the DEVS formalism that enable indirect communication in the spirit of stigmergy. The main difference though is that we enable for multiple hierarchical levels of indirect communication. This difference is significant as this information flows upwards through the model hierarchy propagating upward and enabling information that was not previously possible.

Finally, according to Mittal’s work \cite{Mittal2013} Complex Adaptive Systems (CAS) and its properties are of fundamental interest for the M\&S discipline, with emergent behavior playing a key role in the ladder of complexity. These properties require special attention in order to extend DEVS theory to encompass the CAS theory presenting the following challenges: ``how clusters are formed, hubs appear and evolve; how multi-level self-organization occurs; how strong emergence results in self-organization (with an embedded observer capable of causal behavior at lower levels in the hierarchy); and how formal attention to coupling specification may provide additional abstraction mechanisms to model dynamic interconnected environments'[p.37]. We claim that EB-DEVS capabilities contribute to all these challenges, tackling problems of multi-level modeling, self-organization, strong emergence and stigmergy.


In summary, our work is centered in the modeling and live identification of emergence in the context of DEVS\@. Using the theoretical and conceptual basis discussed so far, we adopt Szabo's definitions of weak and strong emergence, which can be studied with EB-DEVS thanks to its generic, domain-agnostic nature. This strategy allows for flexible definitions of models with emergent behavior within the DEVS realm.

\section{Conclusions and future work}


The main goal of the current study was to determine the feasibility of a DEVS extension focused on the modeling and live-identification of emergent properties in complex systems. In a broad sense, multi-level interactions enable the appearance of emergent properties, facilitating the development of complex systems. For this reason, we emphasize hierarchical models that allow for the generation of system-level properties based on multi-level interactions. 

Centered on the identification of emergent properties, EB-DEVS extended DEVS to generate macro-level `global' states based on micro-level models. Global states can be exposed back to the constituents' state transition functions closing a feedback loop. The implemented bidirectional communication allows for the study of weak and strong emergence.

The use of formal M\&S frameworks for the study of complex dynamic systems offers some warranties while imposing some limitations. On the one hand, it facilitates the modeling and reuse of models while hiding away underlying complexities of the simulation execution. But this comes at the cost of accepting some modeling constraints required by the same formalism. Consequently, we decided to extend DEVS due to its general systems-oriented nature, its modular and hierarchical approach, and its broad acceptance in the M\&S community across several application domains.

To the best of our knowledge, EB-DEVS is the only M\&S framework designed for the live-identification of emergent properties while integrating naturally with existing DEVS-based implementations.

The modeling experience resulting from applying EB-DEVS to three case studies of increasing complexity showed that the mechanisms required to capture emergence remained naturally close to core concepts in complex systems: multiple levels, upward/downward effects, macro/micro states, indirect communication among micro agents and emergent macro properties. EB-DEVS fosters a conceptualization of models in at least two levels (which can in turn be nested in a recursive hierarchy of micro-macro-etc.) producing models that are compact and elegant, which would otherwise be intricate to express. 

Meanwhile, several limitations need to be acknowledged. First, the utilization of Classic DEVS as a base formalism limits the modeling of parallel dynamics. 
Second, the lack of dynamic structural changes limits expressiveness for instance when agents need to be added or removed at simulation time. These aspects will be addressed in future work, together with a systematic exploration of hierarchical feedback loops involving more than two micro-macro levels of abstraction.

\bibliographystyle{unsrt}


\appendix

\gdef\thesection{\Alph{section}} 
\makeatletter
\renewcommand\@seccntformat[1]{Appendix \csname the#1\endcsname.\hspace{0.5em}}
\makeatother

\appendixpage
\addappheadtotoc

\section{EB-DEVS closure under coupling} \label{sec:cuc}

\begin{theorem}{(EB-DEVS closure under coupling.)} \label{theo:cuc}
EB-DEVS models are closed under coupling.
\end{theorem}

The proof relies on the ability of EB-DEVS to express any coupled model, along with its component children models, as a behaviorally equivalent atomic model. We will follow this idea to prove the closure under coupling property in EB-DEVS.

In order to prove that the resultant atomic model ${DEVS}_N$ is equivalent to the coupled model $CN$ we need to define a translation procedure between them.
Let $CN$ be an EB-DEVS coupled model with dependant atomic models ${M_d}$ as defined in \autoref{ssec:formal}.
Let the set of influencers of model $d$ be given by $d \subseteq D \cup \{ CN \}, d \not\in I_d$.

The resultant model will be denoted by

\[{DEVS}_N =\:< X, Y, S, \dint, \dext, ta, \lambda,  Y_{up}, S_{macro}>\]

and otherwise stated we will be referring to its constituent elements.

\subsection{The interfaces}

We define ${DEVS}_N$ to inherit $CN$'s interfaces and communication channels with other models. 
This includes the communication with its siblings in the hierarchy, by means of $CN$'s $X$ and $Y$ sets which are directly inherited from $CN$.
For upward/downward communication with $CN$'s parent, it is sufficient to define 
\begin{equation}
\label{yands}
\begin{array}{rl}
Y_{up}=&Y_{G_{up}} \\ 
S_{macro}=&S_{G_{macro}}
\end{array}
\end{equation}

\subsection{The state set}

S is designed to hold all possible state combinations of the atomic model's states. Thus, $S$ shall hold, for each model, its state $s$, the elapsed time $e$, the $y_{up}$ message sent in the last transition, and the macro level values $s_G$ and $e_G$. The later are needed to compute the macro-level state $s_G$ via the $\delta_G$ function call.

\begin{equation}
\begin{array}{r}
    Q_d  = \{(s_d, y_{up,d}, e_d) | s_d \in S_d, 
      y_{up, d} \in Y_{up, d},  e_d \in [0, ta_d(s_d)]\}
\end{array}
\end{equation}

We define $S$ by taking each possible combination of all model's states in $Q_d$, plus the required global elements of $CN$:

\begin{equation}
\begin{array}{r}
S =  \{(q_d, s_G, e_G) | q_d \in \times_{ d \in D } Q_d , 
 s_G \in S_G , e_G \in \Re_{0}^{+} \}
\end{array}
\end{equation}

The elements $s_G$, $y_{G_{up}}$ and $e_G$ of the coupled EB-DEVS model will be used to compute the corresponding $\delta_G$ global transition function.

\subsection{The time advance function}~\\

This function determines when an atomic model will do its internal transition. The time advance for state $s$ is the minimum remaining time between dependant models.

\begin{equation}
 \sigma_d = ta_d(s_d) - e_d
\end{equation}

\begin{equation}
\begin{array}{rl}
    ta&:S \rightarrow \rm \Re_{0}^{+} \cup \{\infty \} \\
    ta(s)&=minimum \{\sigma_d | d \in D\}
    \end{array}
\end{equation}

\subsection{The set of \textit{imminent} models}~\\

We define it as the set of models with minimum time advance:

\begin{gather}
    \text{IMM(s)} = \{d | d \in D \wedge \sigma_d = ta(s)\} \\
    d^* = \text{Select(IMM(s))}. 
\end{gather}

\subsection{The internal transition function}~\\

This function gives autonomy to the resulting model, and must transition whenever any of the components of $CN$ undergo an internal transition.
It shall resort also to $CN$'s parent information exchanged through their downward/upward channels: $Y_{G_{up}}$ for upward causation and $S_{G_{macro}}$ for downward value coupling. The access to these sets is granted by means of \autoref{yands}.
The $CN$ macro-level state is stored in the $S_{G}$ state set.

The construction of the internal transition shall take into account that atomic models might need to access the coupled model's state $s_G$ to fulfill their state transitions. Let
\begin{equation}
    \begin{array}{rl}
        \dint: S \times S_{macro} \rightarrow S \times Y_{up}
    \end{array}
\end{equation}

The state value of the resulting atomic EB-DEVS can be seen as 
$$s = ((\dots, (s_d, y_{up,d}, e_d), \dots), s_G, e_G)$$, and when transitioning $\dint(s) = s'$. Therefore, $s'$ and $y'_{up}$ are defined as follows.

Consider first that model $d^*$ is the imminent model and $I_{d^*}$ are the influenced models by the output messages of $d^*$. Firstly we define

\begin{equation}
\arraycolsep=0.5pt
\begin{array}{l}
      (s'_d, y'_{up,d}, e'_d) =
      \begin{cases}
          \delta_{int,d}(s_d, v_{down}(s_G)), 0 &\text{if }d=d^*\\
          \delta_{ext,d}(s_d, e_d + ta_{d^*}(s_{d^*}),  & \text{if } d \in I_{d^*} \\
          \quad v_{down}(s_G), x_d), 0 &   \\ 
          s_d, \oslash, e_d + ta_{d^*}(s_{d^*}) &\text{otherwise}
      \end{cases} \\
      \text{with } x_d = Z_{d^*,j}(\lambda_{d^*}(s_{d^*}))
      \end{array}
  \end{equation}

Secondly,

\begin{equation}
    \begin{array}{rl}
        s'_G, y'_{G_{up}} &= \delta_G(s_G, e_G +ta_{d^*}(s_{d^*}), x^b_{micro}, s_{G_{macro}}) \\
        \text{with } x^b_{micro} &= \{ y_{up, d} | y_{up, d} \neq \oslash \} \\
        e'_G &= 0
    \end{array}
\end{equation}

Finally, we can express the resulting pair of values yielded by $\dint$ as follows:

\begin{equation}
    \begin{array}{l}
\dint(s, s_{macro}) =  
 ((\dots, (s'_d, y'_{up,d}, e'_d), \dots), s'_G, 0), y'_{G_{up}}
    \end{array}
\end{equation}

\subsection{The external transition function}~\\

Along the same lines of reasoning presented for the internal transition function, we propose a construction of the external transition function.
Again, it must take into account a potential need to access the coupled model's state $s_G$ to fulfill state transitions. Let

\begin{equation}
\arraycolsep=0.5pt
    \begin{array}{l}
        \dext :  S \times \Re_{0}^{+} \times X \times S_{macro} \rightarrow S \times Y_{up} 
    \end{array}
\end{equation}

The state value and the upward causation value of the resulting atomic EB-DEVS can be seen as \[s = ((\dots, (s_d, y_{up,d}, e_d), \dots), s_G, e_G)\] and \[y_{up}=y_{G_{up}}\] respectively. 

Then the elements that build up $s'$ and $y'_{up}$ are defined as follows:

\begin{equation}
\arraycolsep=0.5pt
    \begin{array}{l}
        (s'_d, y'_{up,d}, e'_d) =
        \begin{cases}
         \delta_{ext,d}(s_d, e_d + ta_{d^*}(s_{d^*}), & \text{if } d \in I_{d^*}   \\ 
         \quad x_d, v_{down}(s_G)), 0 &   \\
         s_d, \oslash, e_d + ta_{d^*}(s_{d^*}) & \text{otherwise}
        \end{cases} \\
    \text{with } x_d = Z_{d^*,j}(\lambda_{d^*}(s_{d^*})) \\
    \end{array}
\end{equation}

and 
\begin{equation}
\arraycolsep=0.5pt
    \begin{array}{ll}
        s'_G, y'_{G_{up}} &= \delta_G(s_G, e_G +ta_{d^*}(s_{d^*}), x^b_{micro}, s_{G_{macro}}) \\
        \text{with } x^b_{micro} &= \{ y_{up, d} \, | \, d \in D, \, y_{up, d} \neq \oslash \} \\
        
        e_G &= 0
    \end{array}
\end{equation}

We can express the resulting pair of values yielded by $\dext$ as follows:

\begin{equation}
  \begin{array}{l}
    \dint(s, s_{macro}) =   ((\dots, (s'_d, y'_{up,d}, e'_d), \dots), s'_G, 0), y'_{G_{up}}
  \end{array}
\end{equation}

Remember the definition of $Z$ from Classic DEVS:

\begin{equation*}
    \begin{array}{l}

      \{Z_{i,j} \, | \, i \in D \cup \{CN\}, j \in I_i \} \\
      Z_{i,j}:
      \begin{cases}
          X_{CN}\rightarrow X_j &\text{if }i={CN}\\
          Y_i\rightarrow Y_{CN} &\text{if }j={CN}\\
          Y_i\rightarrow X_j &\text{otherwise}
      \end{cases}
      \end{array}
  \end{equation*}

This means that all components influenced by the external
input \textit{x} will change their state according to the input $x_d$ transmitted to them converted by the respective interface mappings $Z_{i,j}$. All other components increment their elapsed times by \textit{e}.

Finally, we can express the resulting pair of values yielded by $\dext$ as follows:

\begin{equation}
  \begin{array}{l}
    \dext(s, e, x, s_{macro}) = ((\dots, (s'_d, y'_{up,d}, e'_d), \dots), s'_G, 0), y'_{G_{up}}
  \end{array}
\end{equation}

\subsection{The output function}~\\

In the case of the output function, we define it to dispatch the imminent's output value.

\begin{equation}
    \begin{array}{rll}
    \lambda     &: S\rightarrow Y \\
        \lambda(s) & = Z_{d^*,CN}(\lambda_{d^*}(s_{d^*})) & \text{if } {CN} \in I_{d^*} \\
        & = \oslash & \text{otherwise}
    \end{array}
\end{equation}

To output the corresponding value, we need to translate the output of the selected imminent component
$d^*$ through the interface map $Z_{d^*, N}$.
This is only in the case which $d^*$ has external outputs to send.

We have therefore built the resultant atomic model that represents an initial coupled model and its network of atomic models, concluding the proof.

\section{The abstract simulator} \label{ssec:abstract}

\subsection{The EB-DEVS root coordinator}

The EB-DEVS root-coordinator algorithm is in charge of the initialization of the first level of models, this initialization messages are further propagated by lower level processors.
It also signals via the \textit{*-message} that the first subordinate in transition needs to run.
The \textit{*-message} is modified in order to send the first macro-level state, a \textit{null} value. The higher-level coupled models will have no information from their parents as there is none. In the \autoref{abs:rootcoord} we show in red the modifications that were made to the DEVS abstract simulator.

\begin{lstlisting}[style=base,linewidth=\columnwidth,breaklines=true,
    basicstyle=\ttfamily\footnotesize,
    xleftmargin=2em,
    numbers=left,
    stepnumber=1,
    tabsize=2,
    firstnumber=1,
    numberstyle=\tiny,
    numberfirstline=false,
    label={abs:rootcoord},
    caption={Pseudocode of the EB-DEVS-root-coordinator}]
EB-DEVS-root-coordinator
  // variables:
  t     // current simulation time
  child // direct subordinate devs-simulator or devs-coordinator
  // algorithm:
  t = $t_0$
  send initialization message (i, t) to subordinate
  t = tn of its subordinate
  loop
    send (*, t, @null@) message to child
    t = tn of its child
  until end of simulation
end EB-DEVS-root-coordinator 
    \end{lstlisting}

\subsection{The EB-DEVS-coordinator} \label{ssub:coordinator}

The EB-DEVS-coordinator dispatches messages to other models in the hierarchy and computes the macro-level state when required. The mechanism is detailed in \autoref{lst:coordinator}. 
First, it forwards initialization messages to its dependant models in line 19. Afterwards, it processes four types of messages: initialization messages (i-message), internal transition messages (*-message), external transition messages (x-message) and upward causation messages (y-up-message).

The two main messages handled by the EB-DEVS-coordinator are those in charge of triggering internal and external transitions.

\begin{description}[noitemsep]
   \item[The *-message] is sent to the imminent models that need to run an internal transition. The EB-DEVS-coordinator will forward this message to the minimum remaining time dependants. In EB-DEVS we modified the signature of this message to forward the $v_{down}(s_G)$ value (lines 24 and 28). This is the value-coupling function applied to the macro-level state. After sending this message it will wait for the dependants to raise their $y_{up}$ messages (line 30), queuing them in the input bag $x^b_{micro}$ as they arrive (line 35). After all the messages are received, i.e.\ the ones from all the imminents, and from the influenced models, it will invoke the $\delta_G$ function updating the macro-level state (line 32).
   \item[The x-message] is forwarded towards the processors that need to handle an external transition. These are EB-DEVS-simulators or EB-DEVS-coordinators that will forward the x-message towards other processors. We modified the x-message signature (lines 37 and 42) to include the macro-level state as in the *-message. The main difference is that there is no need to wait for upward causation events, those will be handled in the *-message.
\end{description}

    \begin{lstlisting}[style=base,linewidth=\columnwidth,breaklines=true,
    basicstyle=\ttfamily\footnotesize,
    xleftmargin=2em,
    numbers=left,
    stepnumber=1,
    tabsize=2,
    firstnumber=1,
    numberstyle=\tiny,
    label={lst:coordinator},
    caption={Pseudocode of the EB-DEVS coordinator.},
    numberfirstline=false] 
EB-DEVS-coordinator 
// variables:
CN = (X, Y, D, $\{M_d\}$, $\{I_d\}$, $\{Z_{i,d}\}$, Select, $\color{red}X^b_{micro}$, $\color{red}Y_{Gup}$, $\color{red}S_{Gmacro}$, $\color{red}S_G, v_{down}, \delta_G$) // EB-DEVS coupled model
parent         // parent coordinator
tl             // time of last event
tn             // time of next event
event-list     // list of elements (d, $tn_d$) sorted by $tn_d$ and Select
$d^*$               // selected imminent child
@$\color{red} x^b_{micro}$      // the upward causation set@
$\color{red}y_{Gup}$           @// the mailbox to communicate upwards the new coupled state@
$\color{red}s_{Gmacro}$        @// the parent's state, used as input to transition this coupled model@
$\color{red}s_G$               @// the current coupled model state@
$\color{red}v_{down}$          @// the value coupling function to communicate downwards the new state@
$\color{red} \delta_G$         @// the coupled model's state transition function@
// algorithmm:
when receive i-message(i, t)   at  time t
  for-each d in D do
    send i-message(i, t) to child d
  sort event-list according to $t_n$  and Select
  tl = max {$\mathtt{tl_d}\, | \, d \in$ D}
  tn = min {$\mathtt{tn_d}\, | \, d \in$ D}

when receive *-message(*,  t, $\color{red} s_{Gmacro}$)  at  time t
  if  t != $t_n$ then
     error: bad synchronization
  $d^*$ = first(event-list)
  send *-message(*,  t, $\color{red} v_{down}(s_{G})$) to $d^*$
  sort event-list according to  $tn_d$  and Select
  @wait for all $\color{red} y_{up}$ messages@
  $\color{red} e_G$@ = t - tl@
  @$\color{red} s_G, y_{G_{up}} = \delta_G(s_{Gmacro}, s_G, e_G, x^b_{micro})$@
  tl = t
  tn = min { $tn_d \,|\, d \in$ D}
  @send y-up-message($\color{red}y_{G_{up}}$) to parent coordinator@

when receive x-message(x, t, $\color{red} s_{Gmacro}$) at time t with external input  x and parent state $\color{red} s_{Gmacro}$
  if  not ($tl \leq t \leq tn$)  then
    error: bad synchronization       //consult external input coupling to get children influenced by the input
  receivers = { $r  \, | \, r \in D, N \in I_r,\, Z_{N,r}(x) \neq \oslash$ }
  
  for-each r in receivers
    send x-messages($x_r$, t, $\color{red} v_{down}(s_{G})$) with input value  $x_r = Z_{N,r}(x)$ to r
  sort event-list according to $tn_d$  and Select
  tl = t
  tn = min { $tn_d \, | \, d \in D $ } 

when receive y-message($y_{d^*}$, t) with output $y_{d^*}$ from $d^*$ 
  if  $d^* \in I_N  \wedge  Z_{d^*,N}(y_{d^*}) \neq \oslash$ then
    send y-message($y_N$, t)  with value $y_N = Z_{d^*,N}(y_{d^*})$ to parent
  receivers = { $r \, | \, r \in D, d^* \in I_r, Z_{d^*,r}(y) \neq \oslash $}
  for-each r in receivers
    send x-messages($x_r$, t, $\color{red} v_{down}(s_{G})$) with input value  $x_r  = Z_{d^*,r}(y_{d^*})$ to r

@when receive y-up-message($\color{red}y_{up}$, t) with output $\color{red}y_{up}$ @
  @add $\color{red} y_{up} \text{ to } x^b_{micro}$@
end EB-DEVS-coordinator

\end{lstlisting}

\subsection{The EB-DEVS simulator}
\label{ssub:The-EB-DEVS-simulator}

As detailed in \autoref{lst:simulator}, the simulator works by processing the messages received from the EB-DEVS coordinator. Its main responsibility relies on calling the $\dint$ and $\dext$ events and forwarding the corresponding y-up and y messages to its parent coordinator.
It handles three messages: initialization messages (i-message in line 10), internal transition messages (*-message in line 14), and external transition messages (x-message in line 24).

The following modifications made in the simulator are required to forward the macro-level state into the $\dint$ (lines 14 and 19) and $\dext$ (lines 24 and 28) function invocations. Here we close the loop making available the macro-level information to the atomic models.
After the state transition functions are called, the $y_{up}$ message is forwarded to the processor's parent (lines 22 and 31).

\begin{lstlisting}[style=base,linewidth=\columnwidth,breaklines=true,
    basicstyle=\ttfamily\footnotesize,
    xleftmargin=2em,
    numbers=left,
    stepnumber=1,
    tabsize=2,
    firstnumber=1,
    numberstyle=\tiny,
    numberfirstline=false,
    label={lst:simulator},
    caption={Pseudocode for the EB-DEVS simulator}] 
EB-DEVS-simulator
// variables:
    $M = < X, Y, S, \dint, \dext, ta, \lambda,  \color{red}Y_{up}, S_{macro}\color{black} >$ // EB-DEVS atomic model
    parent    // parent coordinator
    tl        // time of last event
    tn        // time of next event
    y         // current output value of the associated model
    $\color{red} y_{up}$          // output port for communication with the model's parent
// algorithm:
  when receive i-message( i, t)  at time t
    tl = t - e
    tn = tl + ta(s)

  when receive *-message(*,  t, $\color{red} s_{macro}$) at time t
    if  t != tn then
      error: bad synchronization
    y = $\lambda(s)$
    send y-message(y, t) to parent coordinator
    @$\color{red} s, y_{up} = \dint(s, s_{macro})$@
    tl = t
    tn = tl + ta(s)
    @send y-up-message($\color{red}y_{up}$) to parent coordinator@

  when receive x-message(x, t, $\color{red} s_{macro}$) at time t with input value  x
    if not $(tl \leq t \leq tn)$  then
       error: bad synchronization
    e = t - tl
    @$\color{red} s, y_{up} = \dext(s, e, x, \color{red} s_{macro})$ @
    tl = t
    tn = tl + ta(s)
    @send y-up-message($\color{red}y_{up}$) to parent coordinator@
end EB-DEVS-simulator
\end{lstlisting}

\subsection{The abstract simulator execution as a sequence diagram} 

\begin{figure*}[!tbp]
     \centering
     \includegraphics[width=0.9\linewidth]{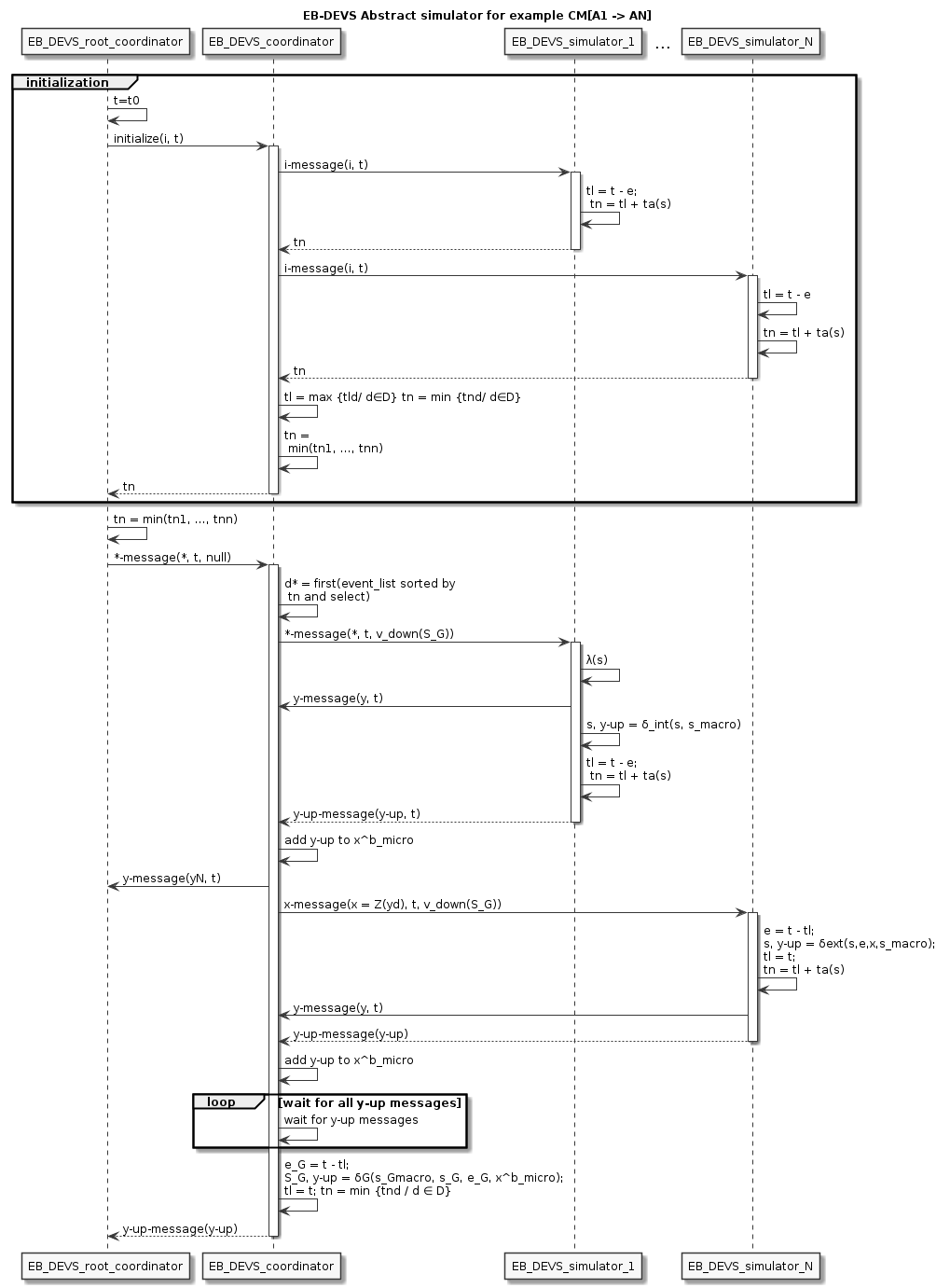}
     \caption{Sequence diagram for the abstract simulator algorithm. Coupled model CM contains A1 to AN atomic models. A1 output port is connected to AN input port. A1 starts its execution with an internal transition, and sends an output message to message to AN generating a cascade of $\delta_G$ model state transition function executions.}
     \label{fig:seqdiag}
 \end{figure*}

In \autoref{fig:seqdiag} we see how the abstract simulator runs for a model comprised of one couple model (CM) with two atomic models (A1, AN). Both atomic models are connected through input-output ports. A1 model will output a message to AN model after the first internal transition execution. In the \autoref{fig:seqdiag} we see how the initialization messages are forwarded from the root-coordinator (the top-level processor) through the coordinator (the internal nodes) towards the leaves (the EB-DEVS-simulator). After the initialization section the EB-DEVS-root-coordinator proceeds with the signaling of the internal transition messages.

Internal transition messages cascade from the root-coordinator towards the EB-DEVS-imulators. After the EB-DEVS-coordinator establishes the next processor to be selected (by selecting the most imminent model), it sends an internal transition message to EB-DEVS-simulator-1.
The EB-DEVS-simulator-1 will invoke the internal transition function of A1 model generating the output corresponding value, cascading a y-message to the EB-DEVS-coordinator. It will then run its internal transition, changing its state and generating an upward-causation y-up-message. The EB-DEVS-coordinator will wait for all the pending y-up-messages as it sends the external transition messages, x-message, to EB-DEVS-simulator-N.

The EB-DEVS-simulator-2 will handle the x-message calling its external transition function so it can change its own state and generate an upward-causation message.

After this step the EB-DEVS-coordinator, having received all the y-up-messages, will compute the new $s_G$  macro-level state.

\section{Bisimulation between DEVS and EB-DEVS}
\label{sec:bisim}

We will show that given an EB-DEVS model it is always possible to construct an \textit{equivalent} DEVS model that simulates its EB-DEVS counterpart, producing the same trajectories of internal states and the same observable traces of input-output events.
We will split the proof in two cases: EB-DEVS models with and without micro-macro interactions.

\textbf{Note:} In this section we will adapt slightly the naming convention for the sake of clarity: $EBD$ and $DEVS$ will denote EB-DEVS and DEVS models, respectively, and subscripts $A$ and $C$ will denote Atomic and Coupled types, respectively. 

\subsection{EB-DEVS without micro-macro dynamics} 

Consider an atomic EB-DEVS model $EBD_A$ where its $Y_{up}$ and $S_{macro}$ elements have been fixed to a \textit{null} value:
\begin{align*}
EBD_A= <X,
        Y,
        S,
        ta,
        \dint,
        \dext,
        \lambda,
        \oslash,
        \oslash>
\end{align*}

$EBD_A$ has therefore no ability to exchange information with its parent.
Such a model can be simulated with an equivalent atomic DEVS model $DEVS_A$:

\begin{align*}
DEVS_A = 
        <X,
        Y,
        S,
        ta,
        \dint,
        \dext,
        \lambda>
\end{align*}

where $X,Y,S,ta$ and $\lambda$ are the same as in $EBD_A$ and \[\dint^{devs}(s)=\dint^{ebd}(s,\oslash)\] and \[\dext^{devs}(s,e,x)=\dext^{ebd}(s,e,x,\oslash)\]

\[\forall s \in S, e \in \Re_{0}^{+}, x \in X\].


As EB-DEVS models are closed under coupling, and so are DEVS models, it follows immediately that a coupled EB-DEVS model $EBD_C$ can be expressed as a coupled DEVS model $DEVS_C$ in the case where $EBD_C$ has no micro-macro interaction with its parent model.

\subsection{EB-DEVS with micro-macro dynamics} 
These models require to express a transformation between models using the notion of bisimulation to show their equivalence. 

We will adopt the notion of \textit{observationally indistinguishable} on its outputs and states, where both models produce the same observations when the same experiments are performed on them by exerting the system through its inputs.
The bisimulation relation between automata can be also referred to as ``system isomorphism'' \cite{Zeigler2018TheoryFoundations}.

According to Rutten's \cite{Rutten1998} bisimulation definition we can specify a relationship $R$ between automata as follows.

\begin{definition}[Bisimulation] 

A bisimulation between two automata $<S, o, t>$ and $<S', o', t'>$ is a relation $R \subseteq S \times S'$ that verifies 

\begin{align*}
    \text{if } s\, R \,  s' \text{ then } \begin{cases}
    o(s) = o(s') \quad \text{ and} \\
    t(s, a) \, R \, t'(s',a)
    \end{cases}
\end{align*}

for all $s \in S, s' \in S'$ and $a \in A$, where $S$ and $S'$ are sets of states, $o$ and $o'$ are output functions, $t$ and $t'$ are transition functions and $A$ is the alphabet of inputs.
\end{definition}

Two automata are bisimilar if there exists a relationship $R$ of bisimulation between them. 

\begin{theorem}{(Bisimulation between EB-DEVS and DEVS)} \label{theo:equiv}
EB-DEVS models are bisimilar with DEVS models.
\end{theorem}

We will show that a relation $R$ exists and can be found for any atomic EB-DEVS model by constructing its corresponding bisimilar atomic DEVS model. Let us set a general enough context for our analysis by considering a coupled EB-DEVS model and its set of atomic EB-DEVS components, all of which make use of their micro-macro dynamics. 

The intuitive idea is to find a transformation procedure that can replace the set of all micro-macro communication channels (between the EB-DEVS parent and its components) with a new, overlaying \textit{broadcast network} of Classic DEVS channels (input/output ports connected with directed links) such that the resulting behavior of each DEVS component is bisimilar to its EB-DEVS counterpart. 

\textbf{Note: }In the limit case where there is only one atomic component, the overlaying broadcast network reduces down to zero ports and links.

We will achieve this by computing a shared, broadcast aggregate state $s_{macro}$ at each dependant model, by embedding a copy of the $\delta_G$ function into each component. Thus, $\delta_G$ will be available for all $\dint$ and $\dext$ transition functions, enabling an "imitation" of a micro-macro closed-loop dynamic, but this time computed locally at each component.

First we want atomic models to be able to broadcast their states via their output function to the rest of components. 

We thus modify the $\lambda$ function of each atomic model, adding the responsibility to communicate its state through a special new broadcast output port, after it has finished sending its (non-broadcast) original messages through its original output ports.

Regarding the external state transition, the modified $\dext$ must split its behavior in two modes: regular and broadcast. When an event comes through the broadcast input port the message value is added to a local input bag and then resumes its previous time advance with $ta=\sigma-e$. This cycle must not affect the $s$ state.
When an event comes through a regular input port, $\dext$ will first compute a local version of the aggregate state $s_{macro}$ by invoking its local replica of $\delta_G$. Then, it will compute the original version of $\dext$ (possibly using the value of $s_{macro}$) to calculate the new $s$ state. In this case, an internal transition must be forced ($ta=0$) with the sole purpose of triggering the $\lambda$ function to broadcast the new value of $s$.

The case for the internal transition function shall also split into two modes, regular and broadcast. The broadcast case is the same as with the $\dext$ function, i.e.\ a forced transition with the purpose of triggering $\lambda$ to send $s$ through the broadcast output port.
In the regular case, the modified $\dint$ will also first compute $s_{macro}$ by invoking  $\delta_G$, and then compute the original $\dint$ (possibly using $s_{macro}$) to calculate the new $s$.

Having explained the nature of the transformation we will proceed to formalize this idea. 

For a given EB-DEVS coupled model ($EBD_C$) we define an equivalent DEVS coupled model ($DEVS_C$). Each model will contain multiple (more than one) atomic components. Atomic models are named $EBD_A$ and $DEVS_A$ respectively.

We leave the case of a model with only one atomic model as a particular case with no interesting properties from the EB-DEVS perspective.

Let the EB-DEVS Coupled model be:
\begin{align*}
    {EBD}_C  = <X_{self}^{ebd}, 
                    Y_{self}^{ebd},
                    D^{ebd}, 
                    \{M_i\}^{ebd},
                    \{I_i\}^{ebd},
                     \{Z_{i, j}\}^{ebd},
                     \text{Select}^{ebd}, 
                     X^b_{micro}, 
                    Y_{G_{up}}^{ebd},  S_{G_{macro}}^{ebd},
                    S_G^{ebd}, 
                    v_{down}^{ebd},
                    \delta_G^{ebd}>
\end{align*}

Let us consider for the sake of simplicity that $EBD_C$ is a root parent (i.e., the topmost in the DEVS hierarchy). Hence $S_{G_{macro}}^{ebd} = Y_{G_{up}}^{ebd} = \{\oslash\}$. The inductive step to prove the general case requires to first flatten a generic coupled model (with arbitrary hierarchical levels) by using the closure under coupling property of $EB-DEVS$, thus obtaining the equivalent root parent used here. This structural induction is straightforward and does not require explicit elaboration.

Let the DEVS Coupled model be:
\begin{align*}
DEVS_C & = <X_{self}^{devs}, 
                Y_{self}^{devs},
                D^{devs}, 
                \{M_i\}^{devs},
                \{I_i\}^{devs}, \{Z_{i, j}\}^{devs},
                \text{Select}^{devs}>
\end{align*}

Let the $EBD_A$ atomic model be:

\begin{align*}
EBD_A & = <X^{ebd},
        Y^{ebd},
        S^{ebd},
        ta^{ebd},
        \dint^{ebd},
        \dext^{ebd},
        \lambda^{ebd},
        Y_{up}^{ebd} ,
        S_{macro}^{ebd}>
\end{align*}

And finally let be the ${DEVS}_A$ Atomic model: 
\begin{align*}
DEVS_A & =
        <X^{devs},
        Y^{devs},
        S^{devs},
        ta^{devs}, 
        \dint^{devs},
        \dext^{devs},
        \lambda^{devs}>
\end{align*}

Consider that the same list of atomic models and priorities are present at each coupled model.

\begin{equation}
    \begin{array}{lr}
        \{M_i\}^{devs} = &  \{M_i\}^{ebd} \\
        D^{devs} =&  D^{ebd} \\
        Select^{devs} =& Select^{ebd}
    \end{array}
\end{equation}

We define the coupled model structure by connecting each atomic model in ${DEVS}_C$ using a fully connected mesh network. This is required to broadcast the atomic states to its siblings.

\begin{equation}
    \begin{array}{l}
        I_i^{devs} = \{ d \in D^{ebd} | d \neq i \}
    \end{array}
\end{equation}

The atomic DEVS model will have the following input, output sets and transformation function:
\begin{equation}
    \begin{array}{rl}
        X^{devs} & = X^{ebd} \cup \{(\text{bIPort}, x) | x \in X^b_{micro} \} \\
        Y^{devs} & = Y^{ebd} \cup \{(\text{bOPort}, y) | y \in Y^{ebd}_{up} \} \\
        \{Z_{i, j}\}^{devs} & = \{Z_{i, j}\}^{ebd}
    \end{array}
\end{equation}

where bOPort stands for Broadcast Output Port and bIPort for Broadcast Input Port. We work with named ports in order to distinguish between the new broadcast ports and the Classic DEVS ports IPort and OPort inherited through $X^{devs}$ and  $Y^{devs}$, respectively.

The new set of atomic model's states is defined by:
\begin{equation}
    \begin{array}{rl}
        S^{devs} &= Y^{ebd}_{up} \times S^{ebd} \times S^{ebd}_{macro} \times broadcast \\
        s^{devs} &= (y_{up}, s, s_{macro}, {broadcast})
    \end{array}
\end{equation}

with $broadcast=\{0,1\}$ acting as a flag to determine when the atomic model is broadcasting a state change to its sibling components within $DEVS_C$. We have  lumped $Y^{ebd}_{up}$ and $S^{ebd}_{macro}$ into the classic $S^{devs}$ state set, i.e.\ the two elements that enable an EB-DEVS model to operate with micro-macro dynamics.

The output function will have two responsibilities. On the one hand it will work as defined in $EBD_A$, and on the other it will output the model's state changes for broadcasting purposes.

\begin{align*}
  \lambda^{devs}: & S^{devs} \rightarrow Y^{devs}  \\
 \lambda^{devs}((y_{up}, s, s_{macro}, {broadcast}))   = &
\begin{cases}
       ({bOPort}, y_{up})  &\quad\text{if broadcast},\\
       \lambda^{ebd}(s) &\quad\text{otherwise.} 
     \end{cases}
\end{align*}

The time advance function will take into account if the model needs to broadcast its state:

\begin{align*}
ta^{devs} : & S^{devs} \rightarrow \Re_{0}^{+} \cup \{+\infty\}  \\
 ta^{devs}((y_{up}, s, s_{macro}, {broadcast})) = & \begin{cases}
       0  &\quad\text{if broadcast,}\\
       ta^{ebd}(s) &\quad\text{otherwise.} 
     \end{cases}
\end{align*}

The internal transition $\dint^{devs}$ function will compute the new state by using the $EBD_A$ internal dynamics. 

\begin{align*}
\dint^{devs} : & S^{devs} \rightarrow S^{devs}  \\
\dint^{devs}((y_{up}, s, s_{macro}, {broadcast})) = & \begin{cases}
        (y_{up} , s , s_{macro}, 0) &\quad\text{if broadcast,}\\ 
        (y_{up}', s', s_{macro}, 1)  &\quad\text{otherwise.}\\
     \end{cases}
\end{align*}

with $(y_{up}', s')=\dint^{ebd}(s,s_{macro})$.

The $\delta^{ebd}_G$ function will be invoked only upon reception of broadcast input messages at bIPort. Therefore, for the external transition function $\dext^{devs}$ we need to consider special cases based on the type of input port.

\begin{align*}
\dext^{devs} : S^{devs} \times \Re_{0}^{+} \times X^{devs} \rightarrow S^{devs} \\
\dext^{devs}((y_{up}, s, s_{macro}, {broadcast}), e, (p, x))  = & \begin{cases}
       (y_{up}, s, s'_{macro}, 0)  &\text{if p=bIPort}, \\
       (y_{up}', s', s_{macro}, 1)  &\quad\text{otherwise.}\\
     \end{cases}
\end{align*}

with $(y_{up}', s')= \dext^{ebd}(s, s_{macro}, e, (p,x))$ and $s'_{macro}=\delta^{ebd}_G(s_{macro}, e, x, \oslash)$.

We have just embedded $\delta^{ebd}_G$ from a coupled $EBD_C$ model into an  atomic $DEVS_A$ model by defining

\begin{align*}
\begin{cases}
 s_G=s_{macro} \\
 e_G=e \\
 x^b_{micro}=x\\
 s_{G_{macro}}=\oslash
\end{cases}
\end{align*}

This concludes the constructive proof.

\section{Legitimacy}
\label{sec:legit}
The legitimacy property of a DEVS model specifies requirements to be imposed for the model to be considered a \textit{system}. Zeigler defined a criterion to assess this property for any DEVS model in \cite{zeigler1976theory}. Namely, the legitimacy property requires that a DEVS model does not undergo an infinite number of state transitions within a finite time interval. Therefore, we will define how and when an EB-DEVS instance can be considered a legitimate model of a system. Intuitively, relying on the bisimulation property, we can first transform an EB-DEVS model into an equivalent legitimate DEVS model, stating that if such transformation exists then the EB-DEVS model is legitimate.

\begin{definition}{(Legitimacy of EB-DEVS models).} \label{theo:legit}
An EB-DEVS model $EBD_A$ is legitimate if and only if a transformation $T$ can be found such that $T(EBD_A) \cong {DEVS_A}$ where $DEVS_A$ is a legitimate DEVS model.
\end{definition}

Consider that $T$ is a transformation built by using \autoref{theo:equiv}. Then if $EBD_A$ and $DEVS_A$ are isomorphic, they will present the same properties. If ${DEVS}_A$ is not legitimate then neither $EBD_A$ can be legitimate. But given that $DEVS_A$ is legitimate under the DEVS legitimacy definition (Zeigler et al.\@, 2018, definition 6.1, p. 158) \cite{Zeigler2018TheoryFoundations} then by construction  $EBD_A$ must be legitimate as well.

\end{document}